# *Topologically protected four-dimensional optical singularities*

**Christina M. Spaegele[1], Michele Tamagnone[1,2,*], Soon Wei Daniel Lim[1], Marcus Ossiander[1], Maryna L. Meretska[1], Federico Capasso[1,*]**

[1] Harvard John A. Paulson School of Engineering and Applied Sciences, Harvard University, Cambridge, MA, USA
[2] Fondazione Istituto Italiano di Tecnologia, Genova, Italy.

*Corresponding Authors

## *Abstract*

Optical singularities play a major role in modern optics and are frequently deployed in structured light, super-resolution microscopy, and holography. While phase singularities are uniquely defined as locations of undefined phase, polarization singularities studied thus far are either partial, i.e., bright points of well-defined polarization, or unstable for small field perturbations. We demonstrate for the first time a complete, topologically protected polarization singularity; it is located in the 4D space spanned by the three spatial dimensions and the wavelength and is created in the focus of a cascaded metasurface-lens system. The field Jacobian plays a key role in the design of such higher-dimensional singularities, which can be extended to multidimensional wave phenomena, and pave the way to novel applications in topological photonics and precision sensing.

## *Introduction*

The field of singular optics explores the wide range of novel effects linked to phase and polarization singularities in electromagnetic fields (*1–7*) and has led to a wide range of applications (*8–11*). The definition of phase singularities is unambiguous: it describes points of vanishing amplitude and undefined phase in a complex scalar field (*2*, *12*, *13*). Examples of phase singularities include Laguerre-Gaussian beams (with azimuthal index $m \neq 0$), which carry orbital angular momentum (OAM) and have lines of zero intensity and undefined phase along their optical axes (*4*).

Polarization singularities in monochromatic fields, on the other hand, have a multivalent definition in literature, requiring only one or more parameters of the polarization ellipse (e.g., azimuthal angle, ellipticity angle) to be singular (*1*, *4*, *8*, *14–22*). They cannot be considered as *complete* polarization singularities as the polarization is either still defined at the singularity (e.g., L lines and bright C points [(*15*), SM]), or are singular only in a specific basis and not topologically protected (e.g., V points, dark C points [(*8*), SM]). They are hence easily destroyed by perturbations arising from many sources, such as stray light and device defects, which have the effect of adding or subtracting complex fields. Such fragility greatly limits their useful application range.

The shortcomings of the polarization singularities investigated so far call for research into the existence and design of complete polarization singularities, i.e., topologically protected points where the polarization is not defined. Polarization patterns have been explored in 2D and 3D configurations (the latter including the phase degree of freedom) in Poincare beams, skyrmions, and unstable singular membranes (*23*, *24*). Our aim here is instead to create a fully topologically protected, complete polarization singularity.

Conversely, one can already find topologically protected phase singularities in random complex scalar fields such as the speckle patterns of polarized monochromatic light reflected by a non-polarizing random medium (*13*, *25–27*). If the speckle pattern is projected onto a two-dimensional screen, several points of vanishing amplitude and undefined phase appear. Sufficiently small perturbations in the field (e.g., by the addition of stray plane waves) do not destroy these phase singularities, but only displace them in space. The stability of these singularities against small field perturbations is guaranteed by the topological structure of wave fields; we call these singular structures *topologically protected* and they are associated with quantized invariant values known as the *topological charge.* The only way to eliminate such singularities is to use a perturbation which is strong enough to bring together topological charges of opposite sign. This overlap will cause these singular structures to annihilate (*13*, *28*).

Here, we show that a topologically protected complete polarization *and* phase singularity can be achieved in the four-dimensional space formed by the three spatial dimensions and the wavelength of light by a direct generalization of the phase singularity protection concept in two dimensions. Such four-dimensional singularities have a well-defined higher-dimensional topological charge. We have realized the 4D singularities using subwavelength-spaced arrays of optical elements (metasurfaces) and probed their topological protection with respect to stray light and device imperfections. The metasurface is designed to create an ellipsoid of light in the focal region of an aspheric lens with a complete and topologically protected polarization singularity at its centre. Spatially-resolved measurements of the polarization in the focal region demonstrate the existence of all possible polarization states around the singularity. We perturb the field at and around the singularity by adding a perturbative polarized field at the singularity position. The singularity is experimentally observed to be topologically protected against such perturbations. Finally, we discuss the applications of this new class of optical fields to STED microscopy, optical metrology, and advancing the fundamental understanding of optical field topology.

## *Generalizing phase singularities*

We will build up to the 4D singularity geometry by first examining the migration of the zeros of a simple 1D function when perturbed. Then we will extend these observations to the 2D phase singularity. Finally, we will generalize the findings to design and characterize the 4D singularity. For all these descriptions, we will only consider the class of infinitely differentiable fields ($\mathcal{C}^\infty$), which is justified as we can write steady-state electromagnetic waves in free space as the linear superposition of a finite number of plane waves. In this paper, regular type indicates scalars, boldface type indicates vectors (e.g., $\boldsymbol{E}$), and overlines (e.g., $\bar{\bar{\boldsymbol{J}}}$) indicate matrices and tensors.

We begin with an arbitrary 1D real-valued function $f: \mathbb{R}^1 \mapsto \mathbb{R}^1$ (Figure 1A). One can classify the zeros of $f$ (i.e., $f(x_0) = 0$) into three categories, assigning them a simple topological charge $m_{1D}$ that is dependent on the function's sign changes when crossing the zero: $m_{1D} = 1$ if the function changes from negative to positive, $m_{1D} = -1$ if it changes from positive to negative and $m_{1D} = 0$ if the function is tangent to the axis at the zero but does not cross it. This topological charge can be formalized as

$$m_{1D} = \lim_{\mu \to 0} \frac{\text{sign}(f(x_0+\mu)) - \text{sign}(f(x_0-\mu))}{2} \qquad (1)$$

As necessary for topological invariants, $m_{1D}$ can be summed across a domain [$x_1,x_2$] to yield information about the domain itself (*29*). This sum is conserved under continuous deformations (i.e., smooth transformation of the domain boundaries), as long as the domain boundaries do not coincide with the zero points.

Suppose an infinitesimal, uniform, real-valued perturbation $\epsilon > 0$ (with $\epsilon \in \mathbb{R}^1$) is added to $f$ (Figure 1B). This perturbation cannot destroy a zero with $m_{1D} = \pm 1$ as one is guaranteed to find a nearby field value which cancels the perturbing field and thus moves the zero to that new position. The only way to destroy a zero with $m_{1D}=\pm 1$ through a uniform perturbation is to increase the perturbation strength to merge and annihilate two zeros of opposite charge, as can be seen from Figure 1A by shifting the function upwards or downwards. On the contrary, zeros with $m_{1D}=0$ (for instance 2nd order zero in Figure 1B) are *not protected* against perturbations: they either immediately annihilate or split into two zero points of opposite charge ($m_{1D}=\pm 1$) (*30*). These $m_{1D}=0$ zero points are hence infinitely rare under experimental conditions. A system can be designed to reach this edge case in theory, but in practice it will not be perfectly realized due to experimental imperfections.

We also notice that for first order zeros (i.e., $f'(x_0) \neq 0$) one can approximate the function by $f(x) = f'(x_0)(x - x_0)$ and the topological charge simplifies to $m_{1D}=\text{sign}(f'(x_0))$ (*31*). That means that for first order zeros the first derivative value is sufficient to fully describe the topological properties of the zero. Uniform infinitesimal perturbations will not destroy the zero but offset it by an amount $\Delta x = -\epsilon/f'(x_0)$. We will see that this is true also in higher dimensions. For higher order zeros ($f'(x_0) = 0$) the protection depends on the detailed behaviour of the higher derivatives.

We summarize these observations by stating that a zero of the 1D real-valued function is topologically protected if its first derivative is non-vanishing at the point, and that this zero is associated with a topologically invariant quantity connected to the sign of the derivative.

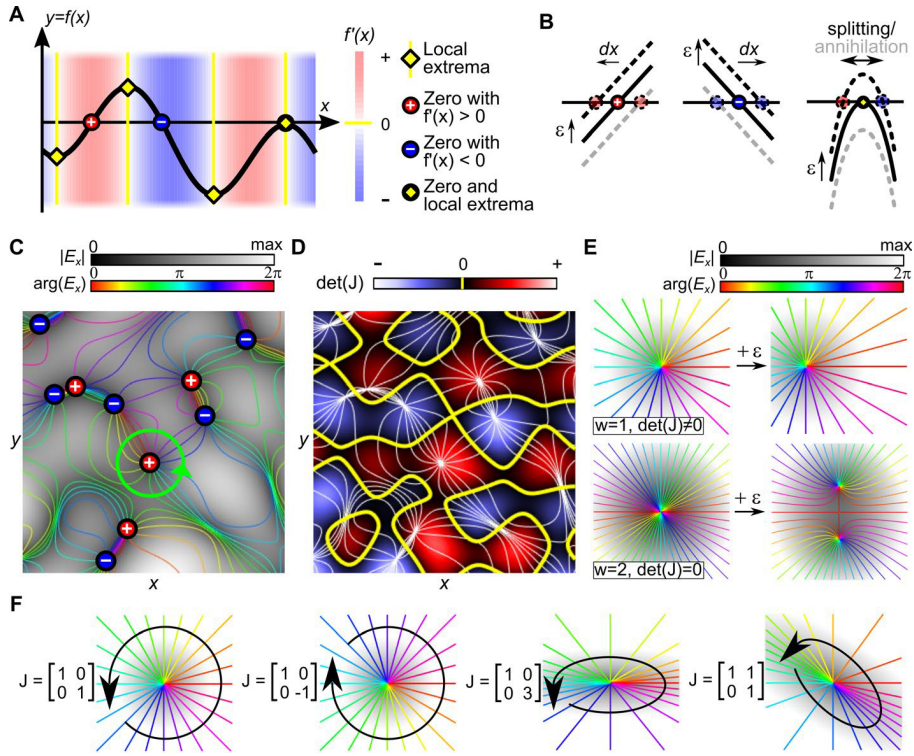

*Figure 1 | Generalization of phase singularities. **A,** A simple 1D function with marked zeros and local extrema. The coloured background represents the derivative value. **B,** Only zeros with a non-zero derivative (topological charge $m_{1D}=\pm 1$) are robust with respect to an infinitesimal perturbation $\epsilon$, which merely displaces them (left and middle panel). The type of zero determines the direction of the shift – if the perturbation shifts the function upwards, zeros with positive derivative move left and vice versa. Zeros that coincide with an extremum ($m_{1D}=0$, right panel) are not topologically protected and either disappear or are split into multiple zeros of opposite charge when the perturbation is added. **C,** Simulation of a random speckle pattern projected on a screen, showing the field's amplitude and phase. Equiphase lines are plotted, which intersect at singularities positioned at points of vanishing amplitude and undefined phase. The phase increases azimuthally from 0 to $2\pi$ around the singularities with a clockwise (counterclockwise) increase in phase corresponding to a negative (positive) topological charge (highlighted respectively). The latter is defined as the winding number $m_{2D}$ and can be determined by the accumulated phase along a closed oriented curve (e.g., along the green loop). Note that only zeros with $m_{2D}=\pm 1$ are found in speckle patterns, corresponding to a $\pm 2\pi$ phase accumulation around the singularity. **D,** Same speckle pattern as in c); here the value of the Jacobian determinant (see text) is plotted showing that each positive singularity falls in a region of positive determinant ($m_{2D}=1$) of the matrix that characterizes the gradient of the complex field (Equation 2) and vice versa. Points where the determinant is null ($m_{2D}=0$) are plotted in yellow. **E,** While singularities with $m_{2D}=\pm 1$ are topologically protected and only displaced by an additive infinitesimal perturbation $\epsilon$ to the field, singularities with $m_{2D}=0$ are destroyed or break into multiple singularities of opposite charges. They are hence not observed in random fields and do not appear in the speckle pattern shown in c). **F,** Amplitude and phase plots of singularities with different orientations, rotations, and skews. The field around the singularity can be described by the Jacobian and does not need to be circularly symmetric.*

The concepts of perturbation protection apply analogously to a 2D complex-valued function $\boldsymbol{E}_x: \mathbb{R}^2 \mapsto \mathbb{R}^2$, and can be used to describe phase singularities in speckle patterns on a 2D screen. Assuming monochromatic light and horizontal polarization, the complex scalar electric field $E_x$ can be represented by its real ($\Re$) and imaginary ($\Im$) parts $\boldsymbol{E}_x = (E_{x\Re}, E_{x\Im})^T$, which depend on the position $\boldsymbol{u} = (x, y)^T$ on the screen (*32*). The speckle pattern shown in Figure 1C is obtained by adding together sinusoids of different spatial frequencies, amplitudes, and phases. Singular positions occur where $|\boldsymbol{E}_x(\boldsymbol{u})| = 0$ with the phase $\phi = \arg(E_{x\Re} + i\,E_{x\Im}) = \mathrm{atan2}(E_{x\Im}, E_{x\Re})$ being undefined at these positions. The phase of the field changes azimuthally around these singularities, that are associated with a positive or negative topological charge depending on their orientation. In 2D, the total topological charge inside a region bounded by a curve C is typically determined as the phase accumulation along C (i.e. $m_{2D} = \frac{1}{2\pi}\oint_C \nabla\phi \cdot d\boldsymbol{l}$), where C by convention is taken counterclockwise (e.g., green curve in Figure 1C) [(*2*), see SM]. This is equivalent to the winding number of the image of C under the function $\boldsymbol{E}_x$, which is a direct generalization of the 1D case. The topological charge of a single singularity is defined taking an infinitesimally small curve C around it.

In analogy to the 1D case, we can write the Taylor expansion around a singularity located at $\boldsymbol{u}_0$ as $\boldsymbol{E}_x(\boldsymbol{u}) = \bar{\bar{J}}(\boldsymbol{u} - \boldsymbol{u}_0)$, replacing the derivative by the Jacobian matrix $\bar{\bar{J}}$, defined as:

$$\bar{\bar{J}} = \begin{pmatrix} \dfrac{\partial E_{x\Re}}{\partial x} & \dfrac{\partial E_{x\Re}}{\partial y} \\ \dfrac{\partial E_{x\Im}}{\partial x} & \dfrac{\partial E_{x\Im}}{\partial y} \end{pmatrix} \tag{2}$$

and considering its determinant. Note that this definition is consistent with the traditional definition of $m = \pm 1$ OAM line phase singularities: close to the singularity, the complex scalar field is approximately $E_x(r, \theta) = re^{\pm i\theta} = r(\cos\theta \pm i\sin\theta) \approx x \pm iy$, up to an overall scale factor. In this case, the Jacobian is:

$$\bar{\bar{J}} = \begin{pmatrix} 1 & 0 \\ 0 & \pm 1 \end{pmatrix} \tag{3}$$

and the determinant equals $\pm 1$.

In direct analogy to the 1D case, there are two cases: for first order zeros, i.e., $\det(\bar{\bar{J}}) \neq 0$, the Jacobian is sufficient to fully describe the topological properties of the zero and the topological invariant becomes $m_{2D} = \mathrm{sign}(\det(\bar{\bar{J}}))$. For higher order zeros, i.e., $\det(\bar{\bar{J}}) = 0$, the topological charge can be determined only examining the higher order derivatives. (*13*, *25*). To better illustrate this for the simulated speckle pattern in Figure 1C, we plot $\det(\bar{\bar{J}})$ in Figure 1D and notice that zeros with positive topological charges are always in a region of positive $\det(\bar{\bar{J}})$, and vice versa.

In analogy to 1D zeros, singularities of this 2D complex field with $m_{2D} = \pm 1$ are topologically protected because they are surrounded by field values of all complex phases (Figure 1E, upper). This means that for an arbitrary small perturbing field $d\boldsymbol{E}$ we can find a nearby point $\boldsymbol{u}'$ in the plane where the field value is $E(\boldsymbol{u}') = -d\boldsymbol{E}$ to cancel out this perturbing field, so that $\boldsymbol{u}'$ is the new singularity location after the perturbation. Higher order singularities, on the other hand, are *not topologically protected* and are either destroyed or split into multiple simpler singularities with $m_{2D} = \pm 1$ by the perturbation (Figure 1E, lower).

The Jacobian plays a critical role in understanding and justifying topological protection of first order zeros. Starting with the Taylor expansion in the vicinity of the singularity, $E_x = \bar{\bar{J}}(u - u_0)$, an additive small perturbation $\varepsilon$ changes the field to

$$E'_x = \bar{\bar{J}}(u - u_0) + \varepsilon = \bar{\bar{J}}(u - u_0 + \bar{\bar{J}}^{-1}\varepsilon) = \bar{\bar{J}}(u - u_0'). \tag{4}$$

For a well-defined, unique $\bar{\bar{J}}^{-1}$, the singularity is hence moved by an amount $\bar{\bar{J}}^{-1}\varepsilon$, and is now positioned at $u_0' = u_0 - \bar{\bar{J}}^{-1}\varepsilon$. Therefore, a singularity is maximally protected (i.e., the singularity position changes the least upon perturbation) if it is surrounded by a large region of a uniform, uniquely invertible Jacobian with large determinant (as $\bar{\bar{J}}^{-1} \sim \frac{1}{\det(\bar{\bar{J}})}$). Note that $\bar{\bar{J}}^{-1}$ is unique and well-defined only if $\bar{\bar{J}}$ is a square matrix and if $\det(\bar{\bar{J}})$ is not zero. A well-defined, unique $\bar{\bar{J}}^{-1}$ further ensures that all phases are found around the singularity. In fact, a small, arbitrary field $dE_x = a(\cos\theta, \sin\theta)^T$ with amplitude $a$ and phase $\theta$ is located at the unique offset $du = -\bar{\bar{J}}^{-1} dE_x$ from the singularity.

Figure 1F shows that the field around the phase singularity changes with the Jacobian and does not need to be circularly symmetric. The orientation, ellipticity, and rotation of the field is determined by the singular value decomposition (SVD) of the Jacobian [SM].

We summarize the connection between topological protection and the non-vanishing of the Jacobian by stating that the topological protection of a singular point of $E_x: \mathbb{R}^2 \mapsto \mathbb{R}^2$ can be ensured if the determinant of the Jacobian is non-vanishing at the point (full proofs in the SM).

These conclusions apply more generally whenever zeros are considered for $f: \mathbb{R}^N \to \mathbb{R}^N$, as the corresponding Jacobian is a square matrix which allows a unique inverse if the determinant is non-zero. While the topological invariant in the 1D case was concerned with the function behavior at points to the left and right of the singularity (more precisely, at the boundary of a 1D interval), and that of the 2D case was based on the function behavior on a closed contour (at the boundary of a 2D surface), the corresponding topological invariants for the N-dimensional case now correspond to function behavior at the boundary of an N-dimensional volume containing the singularity. In the parlance of algebraic topology, this invariant is known as the *topological degree* (*33*) of the function boundary, which is equal to the sum of the topological charges of the singularities inside the volume. For first order singularities, the sign of the determinant is the topological charge. These higher-dimensional invariants are described further in the Supplementary Material.

We can now use the acquired knowledge about the protection mechanism in two dimensions to design a topologically protected polarization singularity. If we consider polarized light beams propagating along a direction *z* within the paraxial approximation (i.e., the z-component of the field is negligible) the electric field vector consists of four real components at each point in space, namely the real and imaginary parts of the *x* and *y* components of the field: $E = (E_{x\Re}, E_{x\Im}, E_{y\Re}, E_{y\Im})^T$. All four components must be zero at the polarization singularity. Again, the Jacobian $\bar{\bar{J}}$ that can be used to describe the field $E(u) = \bar{\bar{J}}(u - u_0)$ can only be uniquely invertible if $\bar{\bar{J}}$ is a square matrix, i.e., we must match the number of parameters to the number of constraints. As $E$ is four-dimensional, it is hence necessary to consider a four-dimensional parameter space $u$ which replaces the 2D screen in the case of the speckle patterns. This can be realized by the three usual spatial dimensions plus the wavelength of light: $u = (x, y, z, \lambda)^T$. The wavelength dependence intended here assumes that the system is illuminated by a light source with tuneable wavelength. As a direct extension of the 2D case, topological charges can again be defined as the degree of the function $E(u) = (E_{x\Re}, E_{x\Im}, E_{y\Re}, E_{y\Im})^T$. A singularity in such a field $E: \mathbb{R}^4 \to \mathbb{R}^4$ from $(x, y, z, \lambda)$ to $(E_{x\Re}, E_{x\Im}, E_{y\Re}, E_{y\Im})$ will be a polarization singularity since both transverse polarizations are undefined ($E_{x\Re} = E_{x\Im} = E_{y\Re} = E_{y\Im} = 0$). We can further engineer the 4x4 Jacobian of this new class of singularities to have a non-zero determinant (and hence a well-defined $\bar{\bar{J}}^{-1}$) to create topologically protected complete singularities that are robust against the addition of arbitrary polarized perturbations. A well-defined $\bar{\bar{J}}^{-1}$ further ensures that all transverse field phases and polarizations exist in the vicinity of this singularity. Analogously to the lower dimensional cases, we can locate a small, arbitrary field $d\boldsymbol{E} \in \mathbb{R}^4$ (and hence any combination of phase and polarization) at an offset $d\boldsymbol{u} = -\bar{\bar{J}}^{-1} d\boldsymbol{E}$ from the singularity position. Hence, the field evaluated at a given distance from the singularity consists of all polarizations and phases, as in skyrmionic hopfions (*23*). In the remainder of this paper, we will show that this new type of singularity can be created and experimentally realized using metasurfaces.

## *Design of 4D optical singularities*

We chose to design the 4D singularity at the centre of a focused light beam (Figure 2A). The light field is generated by a polarization-sensitive metasurface illuminated with horizontally-polarized light [(*34*), SM] and a cascaded aspheric convex lens. The polarization-sensitive metasurface behaves as a spatially varying waveplate and thus gives us control over the local polarization state behind the metasurface with high spatial resolution (*35*). The convex lens relaxes design constraints on the metasurface since the metasurface is then not required to imprint a focusing phase profile.

Our objective is a field distribution around the singularity that can be described by an invertible Jacobian matrix (i.e., $\det \bar{\bar{J}} \neq 0$) to ensure topological protection against perturbations.

We split the design procedure into two phases, starting with the spatial dimensions and later dealing with the wavelength dependence. For simplicity, we initially search for a design where the Jacobian is diagonal in the spatial terms. Let $(dx, dy, dz, d\lambda)^T$ be a displacement from the singularity position in 4D space. First, we consider the spatial structure of the 4D singularity at the design wavelength $d\lambda = 0$. We can describe the field around the singularity as:

$$d\boldsymbol{E} = \begin{pmatrix} dE_{x\Re} \\ dE_{x\Im} \\ dE_{y\Re} \\ dE_{y\Im} \end{pmatrix} = \bar{\bar{J}} \begin{pmatrix} dx \\ dy \\ dz \\ d\lambda \end{pmatrix} = J_0 \begin{pmatrix} 1 & 0 & 0 & J_{14} \\ 0 & 1 & 0 & J_{24} \\ 0 & 0 & 1 & J_{34} \\ J_{41} & J_{42} & J_{43} & J_{44} \end{pmatrix} \begin{pmatrix} dx \\ dy \\ dz \\ 0 \end{pmatrix} = J_0 \begin{pmatrix} dx \\ dy \\ dz \\ 0 \end{pmatrix} \quad (5)$$

Using spherical coordinates around the position of the singularity,

$$\begin{aligned} dx &= dr \sin\theta \cos\phi \\ dy &= dr \sin\theta \sin\phi \\ dz &= dr \cos\theta \end{aligned} \quad (6)$$

this vector $d\boldsymbol{E}$ can also be represented by a complex Jones vector $|d\psi\rangle \in \mathbb{C}^2$ in the horizontal/vertical polarization basis [SM]:

$$|d\psi\rangle = \begin{pmatrix} dE_{x\Re} + idE_{x\Im} \\ dE_{y\Re} + idE_{y\Im} \end{pmatrix} = J_0 \begin{pmatrix} dx + idy \\ dz \end{pmatrix} = J_0[(dx + idy)|H\rangle + dz|V\rangle] = J_0 dr \begin{pmatrix} \sin\theta \, e^{i\phi} \\ \cos\theta \end{pmatrix}, \quad (7)$$

where $|H\rangle = (1,0)^T$ and $|V\rangle = (0,1)^T$ correspond to the horizontal and vertical basis vectors, respectively. The singularity described by equation (7) is surrounded by all possible polarization states at least twice on the sphere (more precisely the ellipsoid) around it (Figure 2B), with points on opposite sides of the singularity having the same polarization but opposite sign (corresponding to a π offset in the phase of the transverse fields).

The required phase and polarization profile can be mapped on the metasurface following the same method used in super-resolution STED (*36*). In absence of the metasurface, the impinging collimated light would constructively interfere at the focal position of the lens. The metasurface acts as a spatially varying waveplate that converts the impinging linearly polarized light into different polarizations (Figure 2C), with the characteristic that each point of the field right after the metasurface has a different polarization except for one counterpart of equal polarization but opposite sign. This leads to destructive interference at the focal position of the lens, since these two polarization states have the same optical path length to the focal position. When one moves away from the focal position, this optical path difference becomes non-zero for certain polarization pairs, leading to incomplete cancellation and no destructive interference. Which polarization pairs are affected depends on the displacement from the focal position, resulting in the polarization distribution shown in Figure 2B [see SM for more details].

The metasurface is implemented by dividing the required profile into square unit cells of periodicity P=420 nm. At each position, we choose the dimension and rotation of a meta-atom that most closely transforms the impinging linear polarized light to the required polarization and phase (i.e. matches the required Jones matrix most closely) at the design wavelength of $\lambda_0 = 600$ nm [(*34*, *35*, *37*, *38*), SM]. We use a meta-atom library consisting of 49613 unique titanium dioxide nanofins of height $h = 600$ nm fabricated on a fused silica substrate.

Up to now, we have ensured that the singularity is surrounded by light (i.e., confined) in the 3D space but we have not yet considered the last dimension $\lambda$. To ensure confinement in $\lambda$, we tune the chromatic dispersion of the metasurface by adding a constant global phase offset to the required Jones matrix profile. This global phase offset can be chosen freely as only differences in the phase across different meta-atoms matter. Changing the global phase over the whole metasurface Jones matrix profile hence does not change the polarization distribution implemented by the metasurface but results in different nanofins with different chromatic dispersions being selected at each meta-atom position [SM].

We emphasize the importance of the dispersion engineering possible with metasurfaces (*39*), as it not only ensures that the Jacobian is invertible, but also enables control of the confinement in the wavelength space. While tuning the global phase is sufficient to create confinement in $\lambda$, dispersion engineering methods described in (*39*) can further improve and shape the confinement in $\lambda$. Implementing the Jones matrix profile selection without this dispersion design may lead to a zero (or very small) determinant of the Jacobian. In that case, a small perturbation could either destroy the singularity (if $\det(\bar{\bar{J}}) = 0$) or move the singularity to a displaced wavelength not reachable by our experimental setup.

Figure 2C show the desired (ideal) and implemented electric field just after the metasurface. The implemented electric field is slightly different from the desired field due to the limited size of the meta-atom library. Simulations of the focal spot profile near the singularity (Figure 2D) that consider the focusing aspheric lens confirm the existence of the complete polarization singularity in the 4D space $(x, y, z, \lambda)$ [see SM].

At the design wavelength, the singularity appears as a single point in the 3 spatial dimensions having null electric field, with all polarizations appearing twice in its immediate vicinity (see Figure 2B) [SM]. As mentioned above, the invertible Jacobian ensures that all polarization *and* phases are located in its immediate vicinity in 4D space *and* ensures its topological protection. The topological charge of the singularity is $m_{4D} = \text{sign}\left(\det(\bar{\bar{J}})\right) = -1$ [SM].

The only way to destroy the singularity by a constant perturbation is to increase it to merge two singularities of opposite sign. For this, one has to push the singularity out of its surrounding region of increasing intensity, which acts like a shield protecting the singularity from destruction. Hence, the singularity is protected as long as the perturbation intensity is smaller than the intensity at the weakest point of this protection shield (i.e., for $0 < I \lesssim I_{max}/2$, with $I_{max}$ being the maximum intensity of the surrounding field). Figure 2E shows the field intensity distribution after an example perturbation of amplitude $\epsilon = \frac{\sqrt{I_{max}}}{3}(1,0,0,1)^T$. While the singularity position is shifted in the four-dimensional space, the minimum intensity value remains constant.

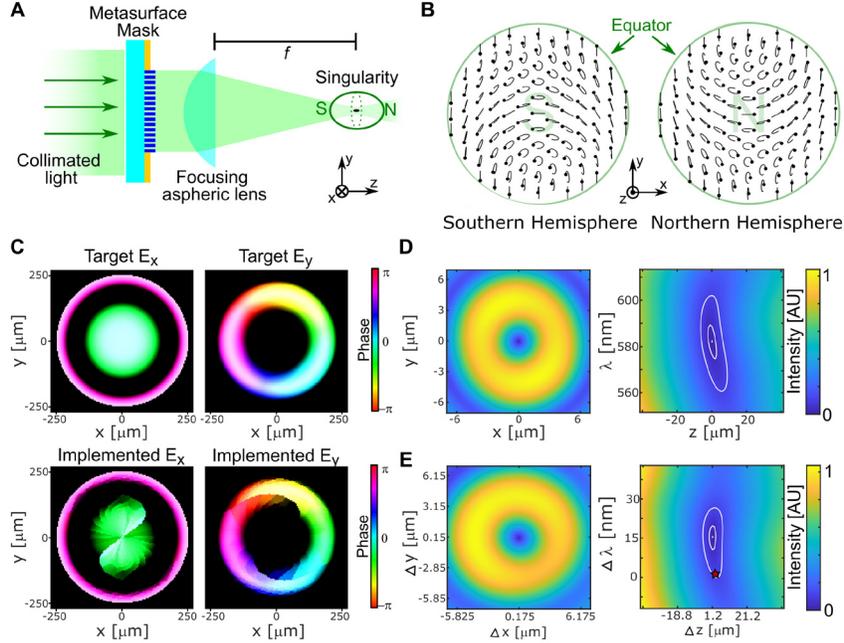

*Figure 2 | Design of 4D singularities. **A,** The singularity is created by shining horizontally polarized collimated laser light through a metasurface that implements the required phase and polarization pattern, and then focusing it with an aspheric lens.* The intensity distribution in the focal region along z is ellipsoidal and centred on the singularity. *The constant light amplitude contour of a cross-section is shown. **B,** The polarization and phase (represented by a dot on the polarization ellipse) of the target electric field around the singularity. All polarizations exist twice on the ellipsoid, with pairs of identical polarization and intensity but opposite phase being positioned on opposite sides of the singularity. This produces destructive interference at the singularity position. **C,** Simulated target and implemented electric field components 1 µm after the metasurface position, where the field's phase and intensity are represented by color and brightness, respectively. For each point in the target field after the metasurface one can find exactly one other point of equal polarization, but with opposite phase. This ensures destructive interference when the electric field is focused by the aspheric lens. **D,** Simulated normalized intensity of the electric field in the xy- and zλ-plane, showing a singularity of null field and its confinement (i.e., of surrounding increasing intensity) in all four dimensions. The contour lines join points of equal field strength. **E,** Simulated normalized intensity of the electric field in the xy and zλ planes when a field perturbation $\epsilon$ is added, which* can *arise from many sources such as stray light and device defects). The coordinate system has its origin at the position of the unperturbed singularity in d). The singularity is shifted in the 4D space. This specific perturbation mostly affects the singularity location in wavelength; the red star marks the position of the unperturbed singularity in the zλ-plane, but the singularity intensity minimum value remains the same.*

## *Experimental validation*

The metasurface (Figure 3B) of diameter d = 500 $\mu m$ was fabricated using the same process as described in [(*40*, *41*),SM] on a fused silica substrate. The metasurface is illuminated by a collimated supercontinuum laser filtered by a tuneable bandpass filter (bandwidth 5 nm) to select visible wavelengths between 485 nm and 700 nm. The quality of the measurement depends on the monochromaticity of the light source, with the intensity measured at the singularity position decreasing with the bandwidth of the laser. An aluminium aperture mask ensures that no light is transmitted outside of the metasurface area. The light is then focused with an aspheric lens of NA=0.08 and imaged through a 75X microscope (Figure 3A). The spatially-varying Stokes polarization state over transverse planes is retrieved using rotating quarter-wave plate polarimetry (*42*). A precise alignment of the components with respect to the optical axis is essential to observe the singularity with high contrast and was achieved with manual and motorized nanopositioners. The full 4D space $(x, y, z, \lambda)$ can be explored by imaging the singularity for different $z$ positions and wavelengths $\lambda$. The field distribution results (Figure 3C-3E) confirm the theory and numerical predictions, showing a confinement of the singularity along all four dimensions; the relative intensity contrast is ~24dB with respect to the intensity maximum, and all polarizations can be found on a small ellipsoid around the singularity (see Figure 3F,G and SM).

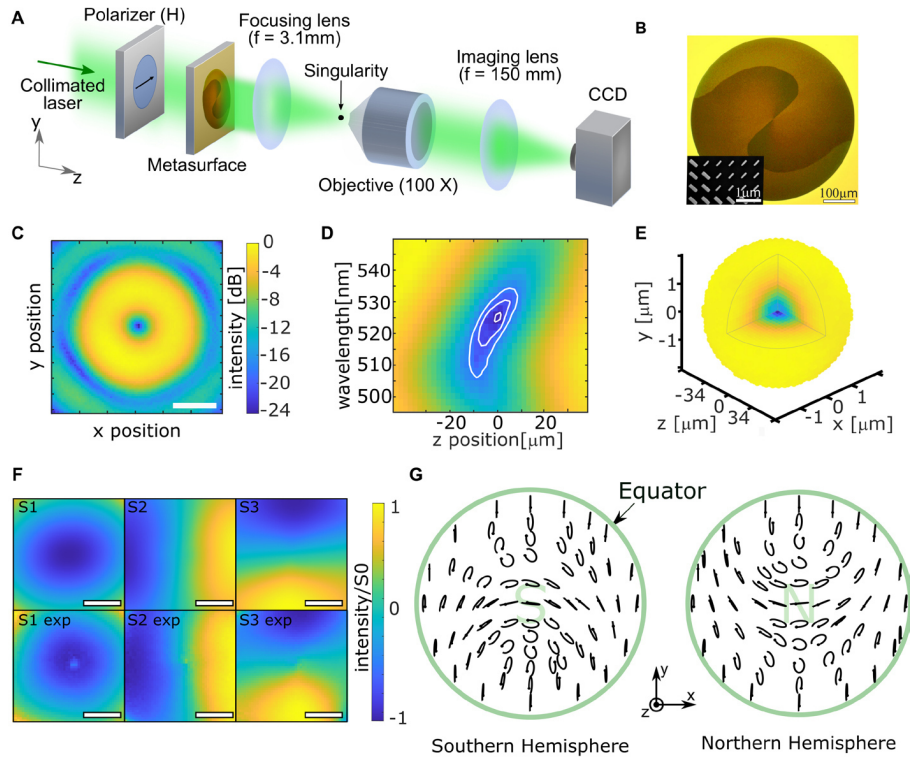

*Figure 3 | Experiment. A,* The experimental setup. The singularity is generated as in Figure 2a and then imaged with a microscope formed by an objective, an imaging lens and an sCMOS camera (which takes pictures of the xy plane). A motorized stage is used to move the objective along the z-direction. The light source is a supercontinuum laser with a tuneable bandpass filter of 5 nm bandwidth, used to explore different wavelengths. To retrieve the polarization distribution in the singularity region, a rotating quarter waveplate and a polarizer were placed into the infinity space between objective and imaging lens [SM]. *B,* Microphotograph of the fabricated metasurface of 500µm diameter surrounded by a gold mask. Inset: SEM image of a small metasurface region. *C-E,* Intensity measurement showing a singularity that is confined in

*4D space: **C,** After determining the position $(x_0, y_0, z_0, \lambda_0)$ of the singularity in the 4D space, an xy image was acquired at $z = z_0, \lambda = \lambda_0$. The dB scale is normalized with respect to the maximum intensity in the 4D dataset. Scale bar 2 µm. **D,** same as C for the z$\lambda$ plane at $x = x_0, y = y_0$. **E,** Sections of the 3D xyz space measured at $\lambda = \lambda_0$ showing that the singularity is fully surrounded by light. **F,** simulated (top) and measured (bottom) Stokes parameters normalized by its pixel-wise intensity $S_0$ at $z = z_0$, $\lambda = \lambda_0$. Scale bar 1 µm, showing good agreement. **G,** Polarization measured on an ellipsoid of equal intensity around the singularity at $\lambda = \lambda_0$, of radius 0.25 µm in the xy plane and length 8 µm in z direction, showing good agreement with the simulation (Figure 2B).*

## *Topological protection*

The fact that the singularity is topologically protected with respect to offsets in the fields provides robustness against perturbation. This also explains why the singularity was easily found experimentally despite imperfections in the metasurface fabrication process and experimental alignment. These imperfections did not destroy the singularity but simply shifted it in space and wavelength. To further observe the topological protection behaviour of our singularity, we insert a small opaque circular gold mask to shadow part of the metasurface. Different areas of the metasurface convert the impinging horizontally polarized light into different polarizations (see Figure 2C and Figure 4A,D) with each polarization having exactly one counterpart of opposite phase on the metasurface, ensuring destructive interference at the focal spot of the aspheric lens. The partial shadowing suppresses the destructive interference of specific polarization pairs at the focus position and hence corresponds to adding the polarized fields with opposite phase at the position of the singularity. Using a gold disk with 110 $\mu m$ diameter, both simulation (Figure 4B,E) and experiment (Figure 4C,F) show that the singularity persists despite the perturbation, while being displaced in the 4D space, i.e., moving to a slightly different position and a slightly different wavelength.

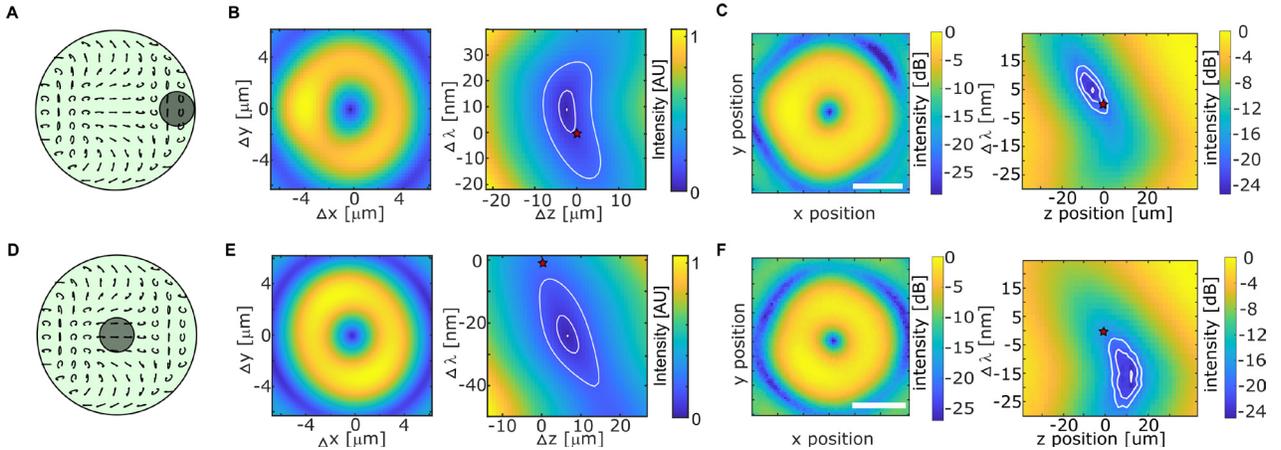

*__Figure 4 | Topological protection.__ **A,** Schematic of the electric field polarization conversion implemented by the metasurface. Blocking part of the metasurface with a gold disk (position and size depicted with light grey circle) corresponds to subtracting a perturbation field of a certain polarization and can be used to probe the protection mechanism of the singularity. **B,** Simulated normalized intensity of the electric field in the xy and z$\lambda$ planes after the perturbation is applied. The singularity is shifted in the z$\lambda$ plane. The red star marks the position of the unperturbed singularity. The minimum intensity of the singularity remains the same upon perturbation. The origin of the coordinate system is the position of the*

*simulated unperturbed singularity (red star marking the position of the unperturbed singularity in the $z\lambda$-plane). **C,** Experimentally measured intensity of the electric field in the xy- and $z\lambda$-plane after the perturbation. The minimum intensity remains the same upon perturbation, again only shifting the singularity in position-wavelength space. The origin of the coordinate system is the position of the experimental unperturbed singularity (red star marking the position of the unperturbed singularity in the $z\lambda$-plane). Scale bar 2 $\mu m$.* **D-F,** *same as A-C, respectively, for a different gold disk position.*

# *Conclusion*

These 4D optical singularities are the direct generalization of phase singularities, and can find applications in light structuring, super-resolution STED microscopy, and polarimetry since they transform the polarization of light into the geometrical displacement of a tightly localized feature. The singularity could also have interesting applications as an exotic dark-field optical tweezer and in ion traps.

The tight localization of the dark spot at the center of the 4D singularity means that the singularity spatial position can be measured with deeply subwavelength precision, better than that of bright regions of light (*43*). This behavior may pave the way for new precision sensors that probe distant physical phenomena (such as position displacements (*44*)) by precisely monitoring minute beam perturbations in the vicinity of the singular region. The polarization and chromatic sensitivity of the 4D singularity expands the palette of detectable parameters that can be deterministically correlated with the 4D singularity position in the combined Cartesian/spectral domain.

More generally, our work paves the way for a new metasurface optics design paradigm, based on engineering not only the structure of light but also its derivatives to achieve fault-tolerant metasurface designs. Such architectures will be ideal for environments with high damage probability, such as in plasma chambers and particle-laden media. Future work will investigate higher and mixed order singularities (which behave with different orders in different spatial directions). Finally, our results are applicable to other design dimensions (for instance the wavelength of light can be replaced by another free parameter of the system, such as the incident tilt angle) and to other wave-like physical systems, as long as they can be represented as smooth maps on real manifolds.

# *Methods*

## *Sample fabrication*

To create the aluminum mask, we spin-coated LOR3A and S1805 resist on a 500 $\mu$m thick SiO$_2$ substrate, exposed it using optical lithography everywhere but at the position of the metasurface, and subsequently developed it in MF319. After an oxygen plasma descum, we deposited 150 nm of aluminum and lifted off by immersing in Remover PG solution for 30 h. Subsequently, we added another 50 nm thick gold mask with a larger opening at its center following the same procedure (for alignment-marker-visibility during electron beam lithography). We then used our standard metasurface process (*40*, *41*) to create the metasurface pillars in the central opening of the mask. (See SM for a detailed process flow).

## Numerical simulations

We created the metasurface pillar library using the rigorous coupled wave analysis solver Reticolo (*45*) and the refractive indexes $n_{SiO2}$ = 1.457 and $n_{TiO2}$ = 2.346. The field calculations around the singularity were performed using Matlab (see SM for a detailed description of the simulations).

## Measurements

The measurements were performed using a supercontinuum laser source. The source generates light in the visible region from 485 nm to 700 nm, which we filter using a tuneable bandpass filter with 5 nm (full width at half maximum) bandwidth. We then focus the light with an aspheric lens (f=3.1 mm, $NA_{eff}$=0.08) and image the singularity using a 100X Nikon objective (NA=0.9), an imaging lens (f=15 cm), and an sCMOS camera (pixel size 6.5 um x 6.5 um, dynamic range 21500:1).

## *Acknowledgements*

The authors thank Noah Rubin for the help with the metasurface library, and Ahmed Dorrah and Joon-Suh Park for the useful discussions. We acknowledge financial support from the Air Force Office of Scientific Research under grants no. FA9550-22-1-024 and FA9550-21-1-0312. M.T. acknowledges the support of the European Research Council (ERC grant no. 948250 SubNanoOptoDevices - ERC-2020-STG) during the last part of the project. This work was performed in part at the Center for Nanoscale Systems (CNS), a member of the National Nanotechnology Coordinated Infrastructure (NNCI), which is supported by the National Science Foundation under NSF award no. 1541959. CNS is part of Harvard University.

## *Author contributions*

M.T. developed the theoretical framework with inputs from C.M.S, S.W.D.L. and F.C.; C.M.S. and M.T. fabricated the devices with help of M.M.; C.M.S., M.T. and S.W.D.L. designed the experiments. C.M.S. performed the experiments with help of M.O.; C.M.S. analyzed the data with help of M.T and S.W.D.L.; C.M.S. and M.T wrote the manuscript with help of S.W.D.L. and inputs from all Authors. F.C. and M.T. led the project.

# *Competing financial interests*

The authors declare no competing financial interests.

# Supplementary Materials for *Topologically protected four-dimensional optical singularities*

## S1.1 Details on polarization singularities in monochromatic paraxial fields

Polarization singularities in monochromatic paraxial vector fields either require only one or more parameters of the polarization ellipse (e.g., azimuthal angle, ellipticity angle) to be singular or are not topologically protected. The sets of points in 2D for which the polarization azimuth is undefined forms C-points, and the sets of points for which the polarization ellipticity angle is undefined forms L-lines (*1*). At these C-points and L-lines, the full transverse polarization can still be well-defined; these singular positions are *not complete* polarization singularities. At C-points, light can be perfectly circularly polarized since circular polarization has an undefined azimuth angle. Similarly, along a L-line, light can be perfectly linearly polarized since linear polarization has an undefined ellipticity angle. Such C-points and L-lines are common in random complex paraxial vector fields (*1–3*). In the main paper, the scalar correspondence of speckle fields has been analyzed in detail. As for phase singularities in speckle patterns, small perturbations in the field (e.g., by the addition of stray plane waves) do not destroy C-points and L-lines, but only displace them. Note that this protection is also guaranteed in 3D nonparaxial fields, where C-points and L-lines turn into C-lines and L-lines, respectively. (The fact that L-lines do not turn into surfaces is described in detail in (*4*) and is based on the fact that for nonparaxial fields 2 conditions need to be fulfilled to ensure an undefined ellipticity angle.).

For V-points and dark C-points, the intensity is zero and the polarization is hence not defined. However, the field in the immediate vicinity of the singularity is polarized in a certain basis (linear and circular, respectively) and will split into multiple bright C-points for, e.g., any elliptically polarized perturbation (*5, 6*). They are hence not topologically protected.

## S1.2 Degree of a function and winding number

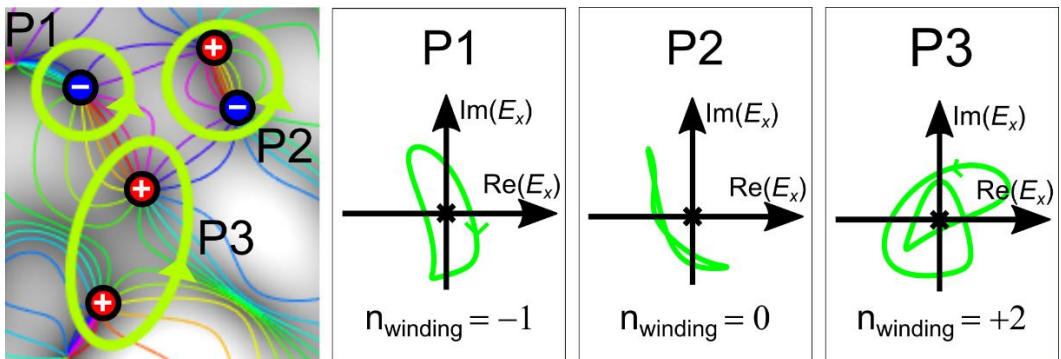

**Figure S1: Winding number and topological charge. a**, Considering a closed path (not self-intersecting or crossing a singularity) on the $(x, y)$ plane (green curves in left panel), we can map the complex field value of each point on this path to a point in the $(E_\Re, E_\Im)$ plane (right panels). The number of times the created closed curve winds around the origin in that plane corresponds to the winding number $w$, which is the sum of the topological charges of the zero points inside the closed path in the $(x, y)$ plane.

The winding number represents the total number of times a closed curve rotates around a point. It is a signed quantity, positive for counter-clockwise rotation, negative for clockwise rotation. The curve can have any shape but must be smooth and not crossing through the point of interest or other singular points. In this paper, the curve lies in the $(E_\Re, E_\Im)$ plane, with $E_\Re$ and $E_\Im$ being the real and imaginary part of the field, and the point of interest is the origin. (Figure S1 left panel). The curve is created by choosing a closed path in the xy plane of the modelled field on a 2D screen (Figure S1 left panel) and plotting the field value at each position of the curve in the $(E_\Re, E_\Im)$ plane (Figure S1 right panel). The closed path in the xy plane is hence not allowed to go through a singularity, as it would correspond to the curve crossing the origin in the $(E_\Re, E_\Im)$ plane. The winding number is then calculated by

$$w = \frac{1}{2\pi} \oint_P \frac{E_\Re \, dE_\Im - E_\Im \, dE_\Re}{E_\Re^2 + E_\Im^2} \tag{S1}$$

Intuitively, this equation integrates over a change in polar angle $\theta = \arctan\left(\frac{E_\Im}{E_\Re}\right)$. As the loop is closed, the overall rotation angle will be a multiple of $2\pi$ so that the field returns to the starting complex value. Under continuous deformation this number is constant because it can be only changed when the loop crosses the singularity point which is not allowed by definition. The winding number is hence a topological invariant under continuous deformations and/or perturbations of the field as long as the curve does not cross a singularity. Formally, this means that for sufficiently small perturbations the winding number is an integer constant.

The field of differential topology provides an immediate generalization for higher dimensions. The winding number is defined for 1D paths in 2D plane $\mathbb{R}^2$, while the degree is valid for the more general case $\mathbb{R}^n$(7). To understand how the generalization works we notice that in the case of the winding number, can see the path $P$ as a smooth function (*diffeomorphism*) from one circumference to another. The first circumference is identified by a parameter $s$ in the range $[0, 2\pi]$ which is used to define the closed path $P$ in the parametric form $P(s)$, such that $P(0) = P(2\pi)$. The second circumference is simply the polar angle (which we assume to be always well defined as discussed earlier) also in the range $[0, 2\pi]$. Then we can identify a function $f: s \to \theta$ and the winding number is simply how many times this function wraps around in $\theta$, with the sign identifying the direction. Practically, this can be achieved with the integral in equation (1) or by taking a regular value $\theta_0$ (i.e., any value for which $f'$ is not zero) and finding all the values $\{s_1, s_2, \ldots\}$ which map to it, so that $f(s_i) = \theta_0$; at each $s_i$ the function can have either positive derivative (counter-clockwise motion) or negative derivative (clockwise motion), and the difference of the number of points for which it is clockwise and the ones for which it is counter-clockwise is the winding number.

Importantly, the points 0 and $2\pi$ have been glued together in both ranges, so that the topology is the non-trivial one of a circumference, also called a *1-sphere*. An *n-sphere* $S^n$ in differential topology is defined as the set of points in a $n+1$ dimensional space which have a distance equal to 1 from the origin. For $n = 1$ it is a circumference in the plane, for $n = 2$ is a spherical surface in the space and so on in higher dimensions. The degree in higher dimensions (in our case we use the case $n = 3$ for a 3-sphere that defines our topological invariant in 4D) is defined similarly to the winding number: starting from the function $f: S^n \to S^n$ we consider a regular value $p$ in the co-domain and the points $\{s_1, s_2, \ldots\}$ which map to it. The derivative is now replaced by the local Jacobian which can be inverting (negative determinant) or non-inverting (positive determinant). The number of points with non-inverting Jacobian minus the number of points with inverting Jacobian is the degree.

The Jacobian is intimately related to this topological invariant: we use here the Jacobian of the function between n-spheres, which depends on the Jacobian of the fields (in the $n+1$ space).

## S1.3 Singular value decomposition of the Jacobian

As described in the main paper, the Jacobian can be used to describe the field near the singularity. For the simplest case of a Laguerre-Gaussian beams (with rotational index m≠0), the Jacobian is the unity matrix (Equation 2), describing a field that is rotationally symmetric in intensity. (Figure S2a). In a more general case, however, the field can be elongated in a certain direction, changing the intensity distribution and the density of the equiphase lines (Figure S2d). The information about the field can be immediately read out of the Jacobian, which can be decomposed into three simple transformations acting on the Laguerre-Gaussian beam profile, using Single Value Decomposition (SVD, Figure S2). Using the SVD, the Jacobian can be written as:

$$J = U\Sigma V^*  \quad (S2)$$

where $V$ corresponds to an initial rotation, $\Sigma$ to a scaling along the coordination axis and U to another rotation.

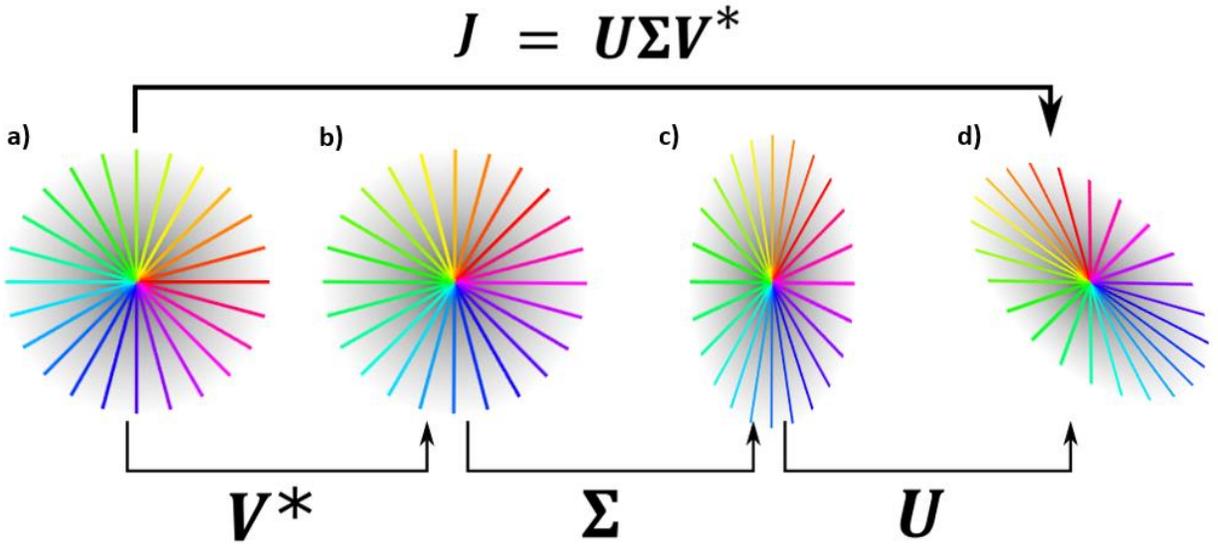

**Figure S2: Singular value decomposition.** The Jacobian $\bar{\bar{J}}$ describes the field around the singularity position. In case of a diagonal unitary Jacobian the field has the shape of an ordinary Laguerre gaussian beam (a). For more general cases, the Jacobian can be decomposed into three simple transformations through Singular Value Decomposition i.e., an initial rotation $V$ (b), a diagonal scaling matrix $\Sigma$ that scales the profile along the coordination axis (c) and a final rotation matrix $U$ (d).

The SVD can be used to express the Jacobian by the parameters of the polarization ellipse, namely, the global phase $\phi$ at the major axis (angle from 0 to $2\pi$), the gradient $a$ along major axis (real positive), the gradient $b$ along minor axis (real, sign is topological charge) and the angle $\theta$ of major axis (angle from 0 to $2\pi$). Then one can write:

$$V^* = \begin{pmatrix} \cos\phi & \sin\phi \\ -\sin\phi & \cos\phi \end{pmatrix}; \quad \Sigma = \begin{pmatrix} a & 0 \\ 0 & b \end{pmatrix}; \quad U = \begin{pmatrix} \cos\theta & -\sin\theta \\ \sin\theta & \cos\theta \end{pmatrix} \quad (S3)$$

and hence

$$M = \begin{pmatrix} a\cos\theta\cos\phi + b\sin\theta\sin\phi & a\cos\theta\sin\phi - b\sin\theta\cos\phi \\ a\sin\theta\cos\phi - b\cos\theta\sin\phi & a\sin\theta\sin\phi + b\cos\theta\cos\phi \end{pmatrix} \quad (S4)$$

## S1.4 Stationary points in the field and the role of the determinant of the Jacobian

In addition to the phase singularities in the 2D spackle pattern case discussed, it has been pointed out [Michael Berry, private communication, 4[th] July 2022, Erice, Italy] that other special points in the field (such as saddle points in the phase) are also relevant to describe the evolution of singularities when the field is perturbed.

Following a more careful analysis, we noticed that the following points all lie on the yellow lines in Figure 2D formed by the points where $\det J = 0$:

- Stationary points (maxima, non-singular minima and saddle points) in intensity
- Stationary points (maxima, minima and saddle points) in the phase
- Stationary points (maxima, minima and saddle points) in the real and imaginary parts of the field

This can be shown mathematically as follows: any real 2X2 matrix $J$ with $\det J = 0$ can be written in the following form parametrized by 3 real parameters $A, B, \theta$:

$$J = \begin{pmatrix} A\cos\theta & B\cos\theta \\ A\sin\theta & B\sin\theta \end{pmatrix} = \begin{pmatrix} \cos\theta \\ \sin\theta \end{pmatrix} \begin{pmatrix} A & B \end{pmatrix} \quad (S1)$$

This implies that for any small displacement in the $xy$ plane, the corresponding offset in the complex plane is $\delta(\cos\theta + i\sin\theta) = \delta e^{i\theta}$ with $\delta$ some real constant, meaning a complex value in the direction $\theta$. If $\theta$ is 0 or $\pi/2$ the point is a stationary point for the imaginary and the real part respectively. Considering now the value of the field in the considered point as $Ce^{i\phi}$, if $\phi$ and $\theta$ are parallel directions, then the point is a stationary point in the phase. If they are orthogonal, then the point is a stationary point in the intensity.

## S2 4D singularity details

## S2.1 Proof that all polarizations exist around the singularity in the 3D space ($dx, dy, dz$) at the design wavelength ($\Delta\lambda = 0$)

Using the inverse function argument in the main paper it is trivial to show that in 4D a neighborhood of the singularity all the polarizations and phases exists. However, we can also prove that in the 3D space (without changing the wavelength) all polarizations exist. Let us define the input space $U$ formed by vectors $u = (\Delta x, \Delta y, \Delta z, \Delta\lambda)^T$ and the output space $V$ formed by vectors $v = (E_{x\Re}, E_{x\Im}, E_{y\Re}, E_{y\Im})^T$

The Jacobian $J$ is full rank since its determinant is non-zero, so spanning around $dx$, $dy$, $dz$ gives three linearly independent vectors in the $V$ space. **It is always possible to combine linearly these three vectors to obtain a $v$ vector with the last two elements set to 0**. This implies that we can always find a point in the 3D space such that the polarization is horizontal. The same argument holds of course for vertical polarization.

For any other arbitrary desired polarization, we can choose a vector $v$ which represents that polarization and has all entries different from zero (using the phase degree of freedom). Then, we can always construct a full-rank matrix $W$ such that $Wv$ is horizontally polarized (i.e., its last two entries are zero). Physically, $W$ could for instance represent a waveplate without losses. Mathematically, a possible construction is:

$$W = \begin{pmatrix} E_{x\Re} & 0 & E_{y\Re} & 0 \\ 0 & E_{x\Im} & 0 & E_{y\Im} \\ -E_{y\Re} & 0 & E_{x\Re} & 0 \\ 0 & -E_{y\Im} & 0 & E_{x\Im} \end{pmatrix}$$

It is easy to verify that the last two entries of $Wv$ vanish, and the matrix is full rank because the determinant is $(E_{x\Re}^2 + E_{y\Re}^2)(E_{x\Im}^2 + E_{y\Im}^2)$ which is greater than zero because no entry is zero.

Let us then consider the matrix $WJ$: we can apply the same argument as above and find a point $u$ in the $U$ space such that the $WJu$ is horizontally polarized, which means that $W^{-1}WJu = Ju$ is the desired polarization. We then conclude that for any desired arbitrary polarization we can find a point in space with that polarization. However, the phase cannot be controlled: only accessing the full 4D $U$ space it is possible to find all the polarizations *and* phases.

This method works because both matrices $J$ and $W$ have full rank. $W$ has full rank because it has no losses, and therefore the product $WJ$ has full rank because the determinant of $WJ$ is the product of determinants which are both non-zero.

## S2.2 Polarization distribution around the 4D singularity

As shown in Figure 2b in the main paper, one can find all polarizations twice on a surface of equal intensity around the singularity. Figure S3. shows the relation between the z position on this surface (a) and the position on the Poincare sphere (b). Figure S3 shows that one can indeed find all polarizations around the singularity, as the Poincare sphere is fully covered when mapping the simulated polarization states of an ellipsoid of constant intensity around the singularity onto the Poincare sphere.

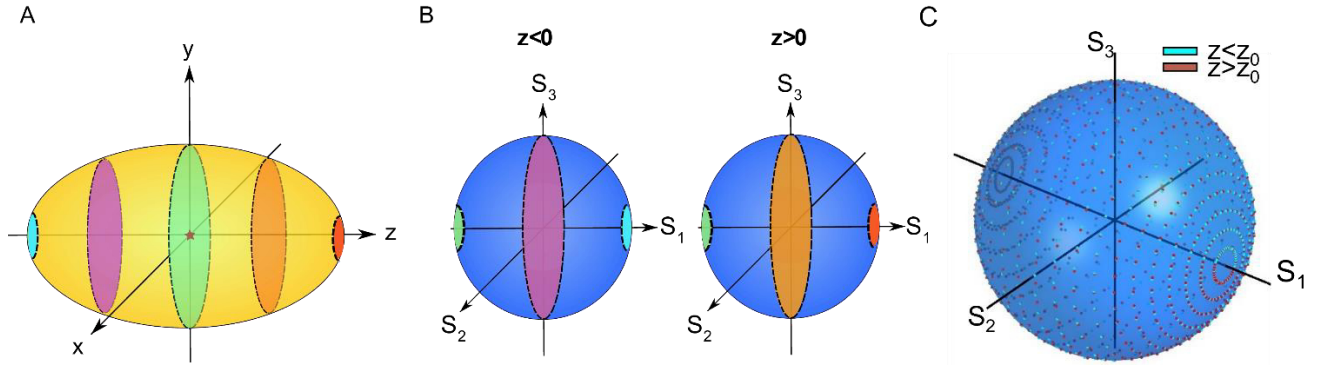

**Figure S3. 3D space mapping to Poincare sphere. a,** Schematic of an surface around the singularity position (star at the origin) of equal intensity. xy-planes located at different z positions are marked in different colors, assuming the singularity is positioned at $(x,y,z,\lambda)=(0,0,0,\lambda_0)$. **b,** Poincare sphere for z larger and smaller than zero. It shows that each point on the Poincare sphere is crossed twice when the polarization states on the surface in a) are mapped onto the Poincare sphere. **c,** Simulated polarization on ellipsoid surface (axis $l_x = l_y = 0.3\,\mu m$, $l_z = 10\,\mu m$) of equal intensity plotted on the Poincare sphere.

## S3 Metasurface design and simulation details
## S3.1 Change of basis between Equations 4 and 6
Starting from Equation 4 in the main paper

$$dE = \begin{pmatrix} E_{x\Re} \\ E_{x\Im} \\ E_{y\Re} \\ E_{y\Im} \end{pmatrix} = \bar{\bar{J}} \begin{pmatrix} dx \\ dy \\ dz \\ d\lambda \end{pmatrix} = J_0 \begin{pmatrix} 1 & 0 & 0 & J_{14} \\ 0 & 1 & 0 & J_{24} \\ 0 & 0 & 1 & J_{34} \\ J_{41} & J_{42} & J_{43} & J_{44} \end{pmatrix} \begin{pmatrix} dx \\ dy \\ dz \\ 0 \end{pmatrix} = J_0 \begin{pmatrix} dx \\ dy \\ dz \\ 0 \end{pmatrix}$$

we can express the field as a Jones vector

$$|d\psi\rangle = \begin{pmatrix} E_{x\Re} + iE_{x\Im} \\ E_{y\Re} + iE_{y\Im} \end{pmatrix} = J_0 \begin{pmatrix} dx + idy \\ dz \end{pmatrix} = J_0 \left[ (dx + idy)\begin{pmatrix}1\\0\end{pmatrix} + dz\begin{pmatrix}0\\1\end{pmatrix} \right] = J_0[(dx+idy)|H\rangle + dz|V\rangle]$$

## S3.2 Metasurface design
The goal of the system composed by the metasurface and the lens is to create a focused beam of light hosting the 4D singularity at its focal point. The key idea is to use the Green's function approach to compute the contribution of each region of the metasurface to the electric field in the focal point of the lens and in its neighborhood. This can be done analytically using a few assumptions about the system, which are satisfied by the experimental system. First, we will use the paraxial approximation to describe the beam after the lens. Second, all the focusing is performed by the lens, while the metasurface simply implements the required phase and polarization profile. It is possible to show that using a lossless metasurfaces based on rectangular pillars it is always possible to obtain a desired transmitted polarization and phase (represented by a Jones vector) given an input polarization and phase (8). In short, this is because any polarization can be converted to another by a proper wave plate, and an additional global delay can control the phase. In practice, the coverage is usually slightly lower than 100%, but this does not affect the formation of the singularity thanks to the fact that it is topologically protected.

At the focal point of the lens, we can deduce that the electric field is given by the integral of all the fields contributions over the metasurface area. To ensure that the field is zero, we design the metasurface to produce pairs of polarizations with opposite signs (phase shift of $\pi$), so that all contributions sum to zero at the focal point (Fig S4A). Away from the focal point, the sum is not vanishing because of the additional phase delays introduced by the offset in the position. This idea is used routinely in other applications requiring 3D holography, including the generation of deexcitation beams for superresolution STED (9) and is summarized here.

- An offset dx with respect to the focal point is equivalent to a phase advance of the left side of the metasurface and a phase delay on the right side (or vice versa), Fig S4B.
- An offset dy with respect to the focal point is equivalent to a phase advance of the top side of the metasurface and a phase delay on the bottom side (or vice versa), Fig S4C.
- An offset dz is equivalent to a certain phase delay (or advance) in the center of the metasurface and a smaller phase delay (or advance) on the rim of the metasurface. Normalizing all fields with the average phase (which can always be done without affecting the continuity of the fields), this is equivalent to a phase advance at the center of the metasurface and a phase delay on the rim (or vice versa), Fig S4D.

Because the target polarization and phase around the singularity is $|\psi\rangle = (dx + idy)|H\rangle + dz|V\rangle$ this is equivalent to mapping the vertical polarization at the center ($r = 0$) and at the rim ($r = r_0$) of the metasurface, and the horizontal polarization in a circle at $r = \sqrt{0.5}r_0$. This factor is chosen to ensure that the rim region ($r > \sqrt{0.5}r_0$) has the same area of the center region ($r < \sqrt{0.5}r_0$) to balance the sum to 0 at the focal point. Additionally, an OAM-like azimuthal phase profile has to be imparted on the horizontal polarization.

These considerations provide the ansatz that the mapping can be performed in a one-to-one manner from the metasurface to the region of space around the focal point (Fig S4E), choosing the desired Jones vector to be:

$$\begin{pmatrix} E_x \\ E_y \end{pmatrix} = e^{i\phi_0 + i\theta} \begin{pmatrix} \cos\left(\frac{\pi r}{r_0}\right)^{2+\varepsilon} \\ \sin\left(\frac{\pi r}{r_0}\right)^{2+\varepsilon} \end{pmatrix}$$

where $r, \theta$ are the polar coordinates on the metasurface, $r_0$ is the radius of the metasurface, $\phi_0$ is a global phase factor and $\varepsilon$ is a small correction to the exponent. The ansatz is then validated by the simulations, which show that the desired profile is obtained around the focal point

To ensure the correct behavior with the wavelength, several free parameters were used:
- The input polarization
- The correction $\varepsilon$
- The global phase.

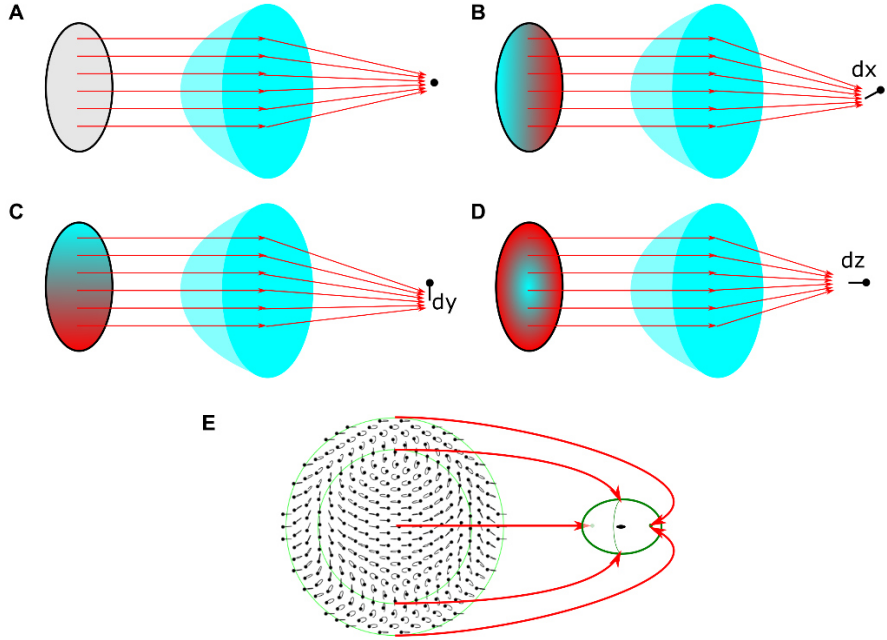

**Figure S4: Design of the phase and polarization profile.** A, at the focal point the intensity of light is zero because polarizations cancel each other in pair. B-D, offsets in the 3D space are equivalent to additional phase gradients on the metasurface. E, mapping from the metasurface to the fields around the singularity.

## S3.3 Metasurface library simulation and metasurface implementation

The design principle of using rotated rectangular pillars to fully control phase and polarization of light is described in (8). The metasurface library is composed of rectangular pillars of height $L_z = 600nm$ and varying length and width and is depicted in Figure S5. The unit cell size was chosen to be $U_x = U_y = 420nm$. The phase delay of the meta atom was simulated using the RCWA-software Reticolo (10), assuming $n_{SiO_2} = 1.46$ and $n_{TiO_2} = 2.4$.

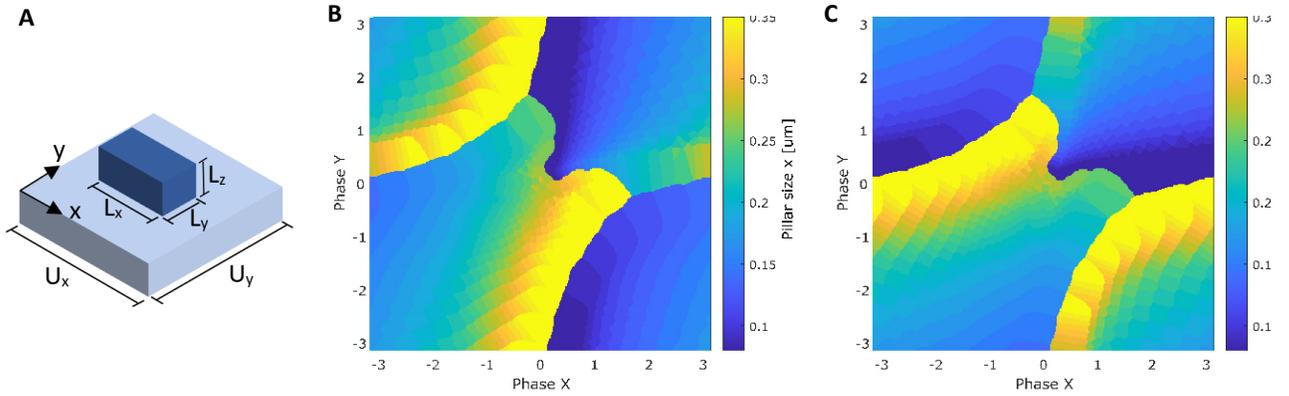

**Figure S5: Metasurface pillar library. a,** Depiction of a unit cell design. A rectangular TiO$_2$ pillar ($n_{TiO2}$ = 2.4 of varying length and width but constant height L$_z$= 600nm is placed in the center of a unit cell of dimension U$_x$=420nm, U$_y$=420nm. The horizontally polarized light (E$_x$) impinges from the side of the SiO$_2$ substrate ($n_{SiO2}$ = 1.46, simulated as semi-infinite). **b,** Direction dependent phase delay for a single wavelength (600nm) vs Pillar dimension L$_y$. **c,** Direction dependent phase delay for a single wavelength (600nm) vs Pillar dimension L$_x$.

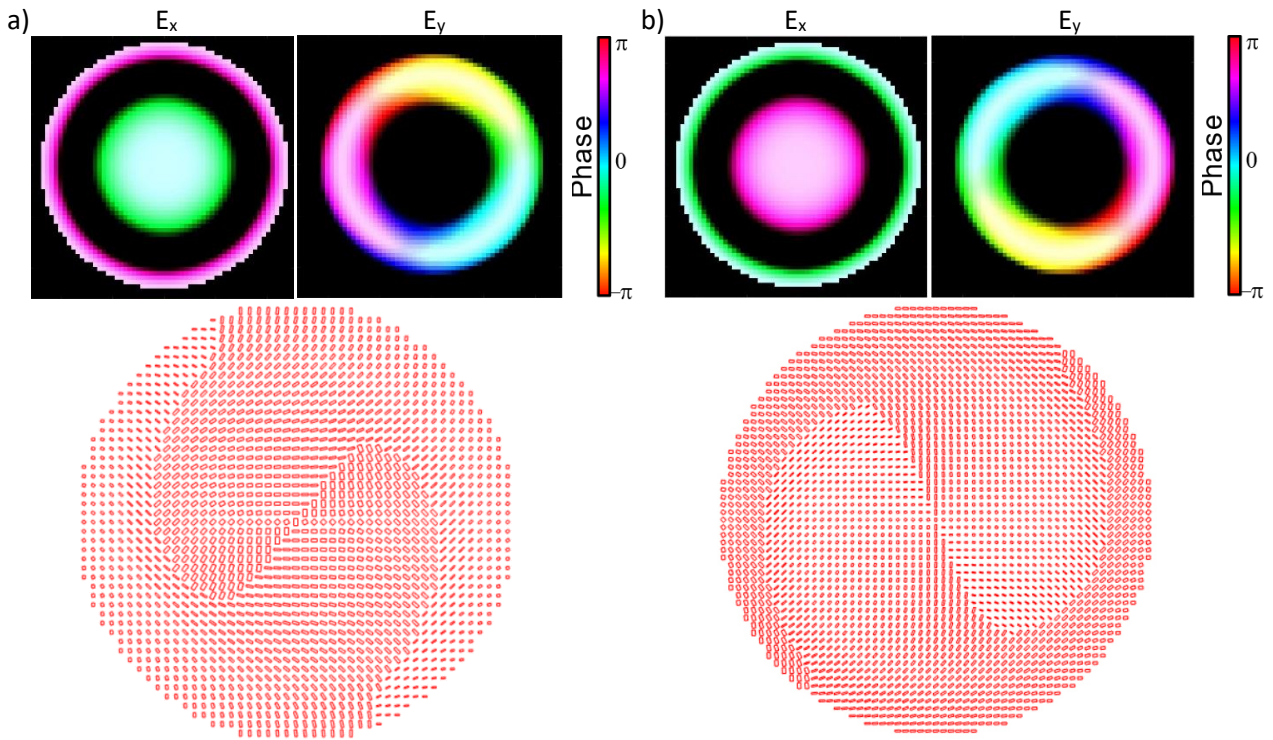

**Figure S6: Metasurface implementation and global phase.** Schematic of the metasurface implementation, sampling every 25th pillar. **a,** metasurface used for the experiment with global phase $\phi_{global} = -0.25\pi$. **b,** same phase profile implementation, but with global phase $\phi_{global} = 0.75\pi$, that implements the same polarization, but changes the selection of pillars.

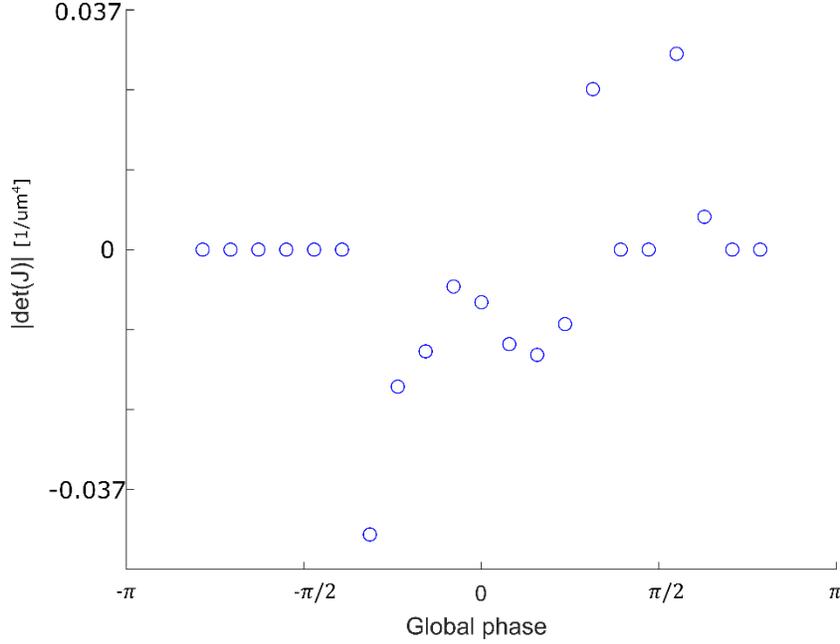

**Figure S10. Influence of the global phase on the Jacobian.** Global phase against the Jacobian determinant at the position of the singularity, showing that the global phase can control the wavelength confinement and can be used to ensure |det(J)|>0.

### S3.4 Simulation of the electric field

The field around the singularity is simulated using a green function integral:

$$E(x_s, y_s, z_s, \lambda_s) \sim \iiint_{MS} dx\, dy\, dz\, \mathrm{T_{MS}}(x, y, z, \lambda)\, \underbrace{e^{-i\frac{2\pi}{\lambda_s}\sqrt{x^2+y^2+f_{lens}^2}}}_{\text{Aspheric lens}}\, \underbrace{e^{i\frac{2\pi}{\lambda_s}\sqrt{(x-x_s)^2+(y-y_s)^2+z_s^2}}}_{\text{Green function}} \quad \text{(SX)}$$

where $f_{lens} = 3.1\text{mm}$ is the focal length of the asphere and $\mathrm{T_{MS}}$ corresponds to the complex transmission right after the metasurface sampled with nanostructures from the library described in S3.2. Note that the Green's function is approximated by discarding the inverse square of the radius decay, which has negligible variation. To reduce computation time, the metasurface was assumed to be ten times smaller than the true size. However, this does not change the resulting normalized field distribution, and it was verified using different scaling values and always obtaining the same results. As the metasurface and the aspheric lens are placed close together, the diffraction effects from the metasurface edges are neglected. To simulate the effect of a perturbation shadowing part of the metasurface, the corresponding part of the metasurfaces transmission profile is set to zero.

As the phase profile of the sampled metasurface does not perfectly match the ideal phase profile (Figure 2c), the singularity will not be placed perfectly at the design position $(x_0, y_0, z_0, \lambda_0) = (0, 0, 3.1mm, 600nm)$, but will be slightly displaced in 4D space even in the simulation. We hence use a root finding algorithm to find the singularity. Specifically, we use the iterative Newton Raphson algorithm following:

$$u_{n+1} = u_n - f(u_n) * J^{-1}(u_n)$$

with $u_n = (x_n, y_n, z_n, \lambda_n)$ being the position in 4D space for the nth step, $f(u_n)$ being the complex field value at position $u_n$ and $J^{-1}(u_n)$ being the inverted Jacobian at position $u_n$. Starting from different positions around the design position, we see a convergence to the singularity that is indeed shifted in the 4D space (Figure S7).

This algorithm also ensures that the simulated singularity is a first order and not a higher order zero in four dimensions. Choosing the starting position u₀ randomly around the found singularity, we can observe its convergence behaviour (Figure S8). While this algorithm converges for first order zeros, it would not converge for higher order zeros as $J^{-1}$ diverges in at least one entry the closer the walker gets to higher order zeros. When divergence occurs, the sequence of points jumps chaotically instead of converging, and it is worth mentioning that the Newton's fractal is related to the convergence/divergence pattern of this algorithm.

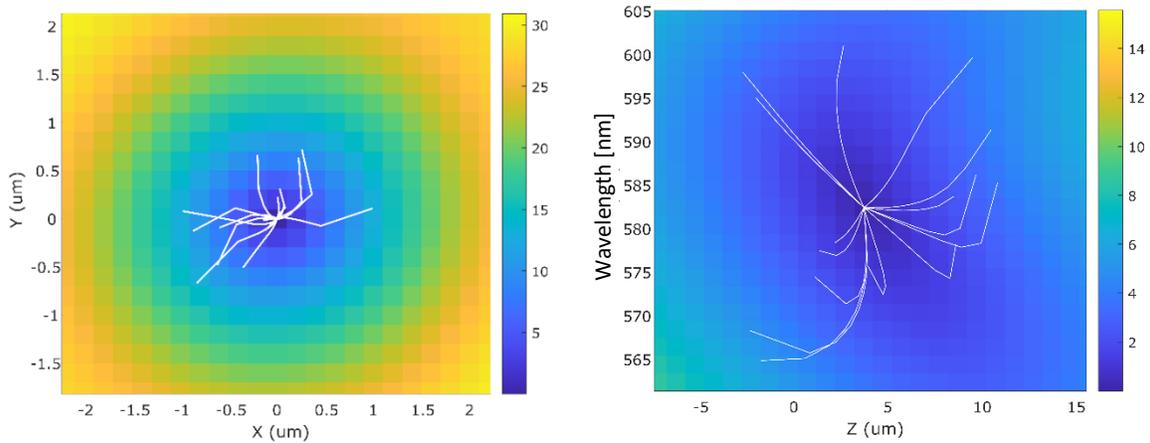

**Figure S7. Finding the singularity using the Newton Raphson algorithm.** The paths converge to the true singularity position.

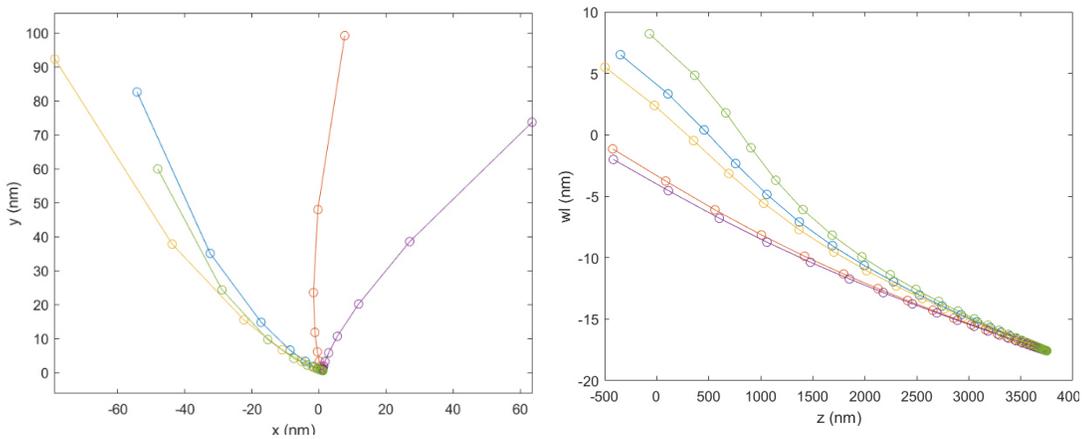

**Figure S8. Validation of 4D singularities: The Newton Raphson algorithm.** Simulations show a singularity in space and wavelength with trajectories (white) of the Newton-Raphson method.

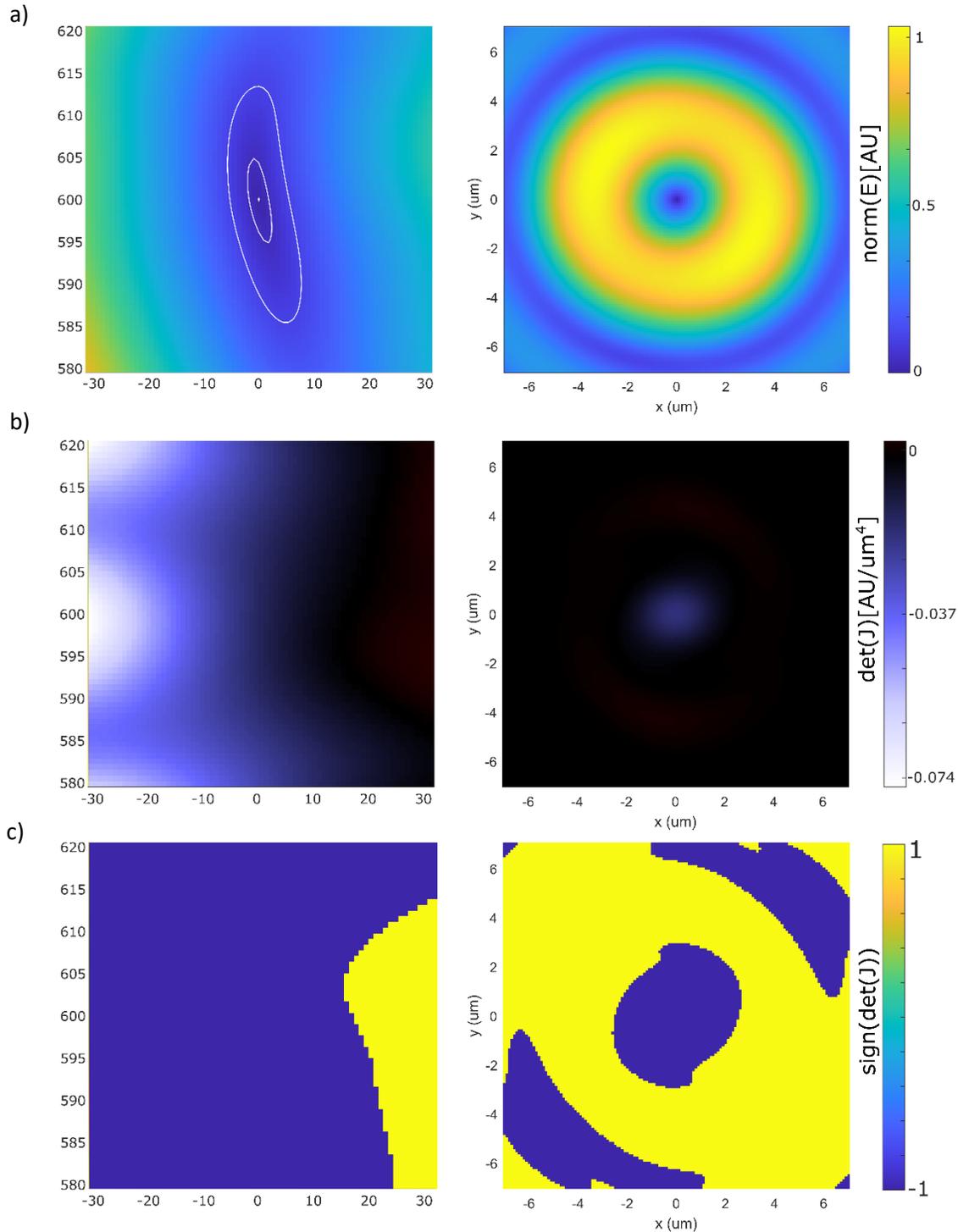

**Figure S9. Validation of 4D singularities: Calculating the Jacobian. a,** normalized field around the singularity in 4D. **b,** corresponding Jacobian determinant in 4D showing that the singularity is positioned in a region of negative Jacobian hence $m_{4D} = \text{sign}\left(\det(\bar{\bar{J}})\right) = -1$. **c,** corresponding sign of the Jacobian determinant.

# S4 Metasurface fabrication details

To ensure that light can only pass through the metasurface and is blocked otherwise, the metasurface is fabricated into an open aperture of an Aluminum mask. The steps for the mask fabrication are summarized in Supplementary Figure S11. To define the position of the hole and the alignment markers, LOR3A and S1813 resist were spin coated on a glass substrate, exposed with optical lithography (maskless aligner) and developed with MF319. 150nm of Al were then deposited in a vacuum E-beam Evaporator and the resist was removed using Remover PG at 80 C. The same procedure was repeated using 50nm of gold to make the alignment markers visible in the electron beam lithography for the following metasurface writing. The diameter of the gold mask hole was chosen to be larger than the metasurface as it would further increase the resist height close to the mask boundary due to capillary forces and hence complicate the metasurface fabrication.

Supplementary Figure S12 depicts the subsequent metasurface fabrication in the center of the mask opening. 600nm of ZEP resist were spin coated onto the mask, exposed with electron beam lithography, and developed with cold o-xylene. TiO2 was deposited via ALD on the patterned resist, and the excess oxide was etched back using a fluorine based RIE recipe. The resist was removed in Remover PG at 80 C. The sample was rinsed in acetone, IPA, and cleaned using oxygen plasma.

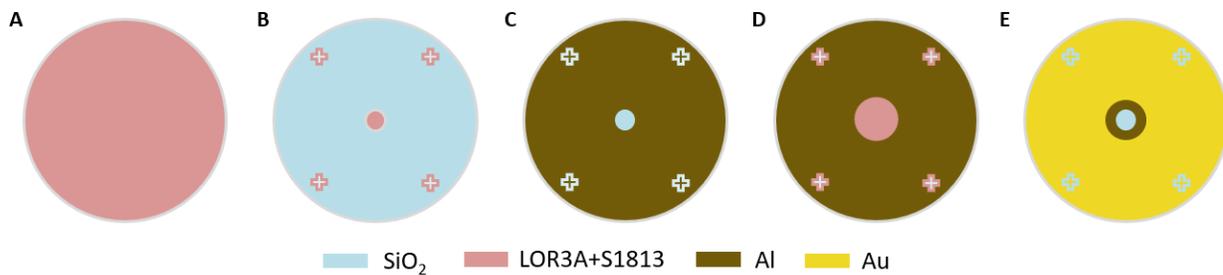

Figure S11. Aluminium and gold mask fabrication. a, Spin coating optical lithography resist. b, Optical lithography exposure and development freeing everything but alignment markers and the final hole of the mask. c, Al deposition using Vacuum E-beam Evaporator and lift-off in remover PG. d, Spin coating optical lithography resist, optical lithography exposure and development freeing everything but alignment markers and a region larger than the initial hole in the Al mask. e, Gold deposition using Vacuum E-beam Evaporator and lift-off in remover PG.

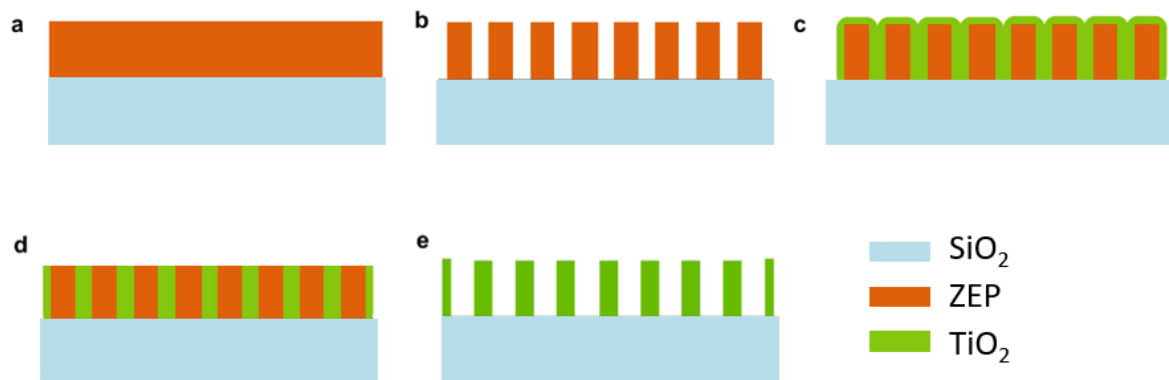

**Figure S12. Metasurface fabrication. a,** Spin coating e-beam resist. **b,** e-beam exposure and development. **c,** TiO2 deposition using ALD. **d,** RIE etch back. **e,** Resist removal and final cleaning.

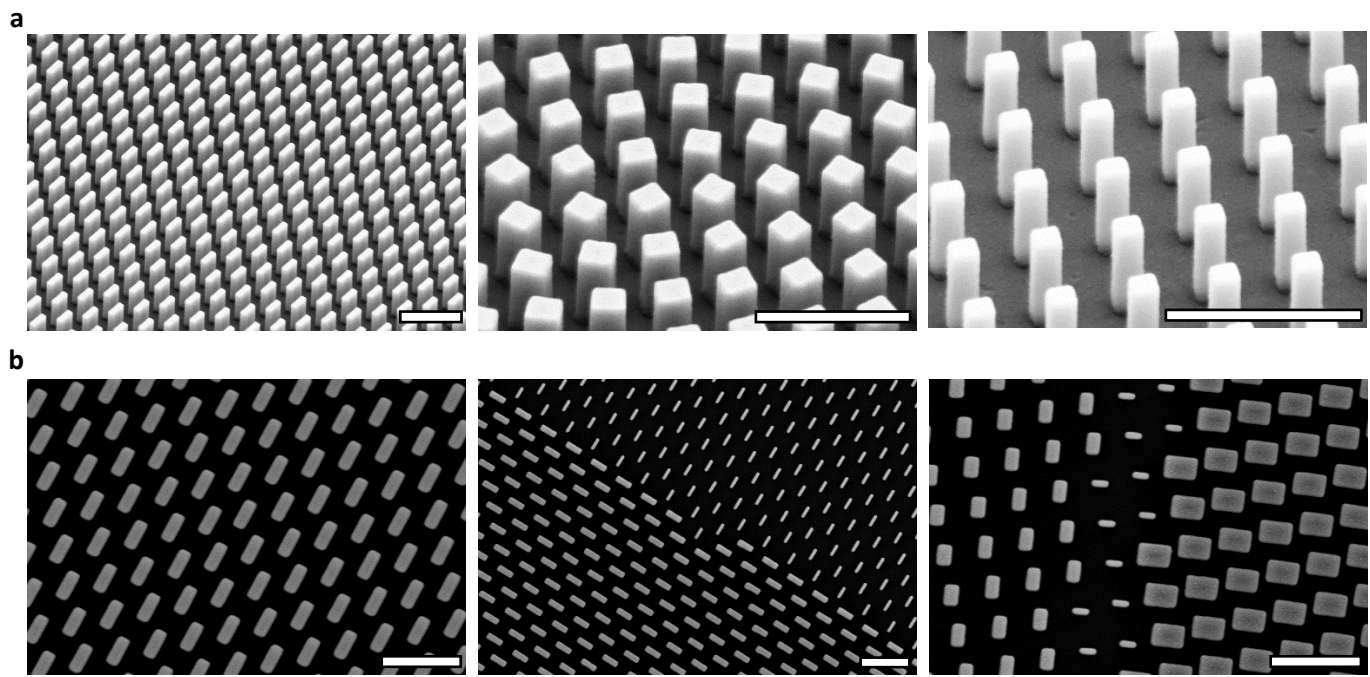

**Figure S12. SEM images of the fabricated metasurface. a,** Different regions of the metasurface. Images taken under an 40° angle. Scale bar $1\mu m$. **b,** Different regions of the metasurface (top view). Scale bar $1\mu m$.

## S5 Measurement details

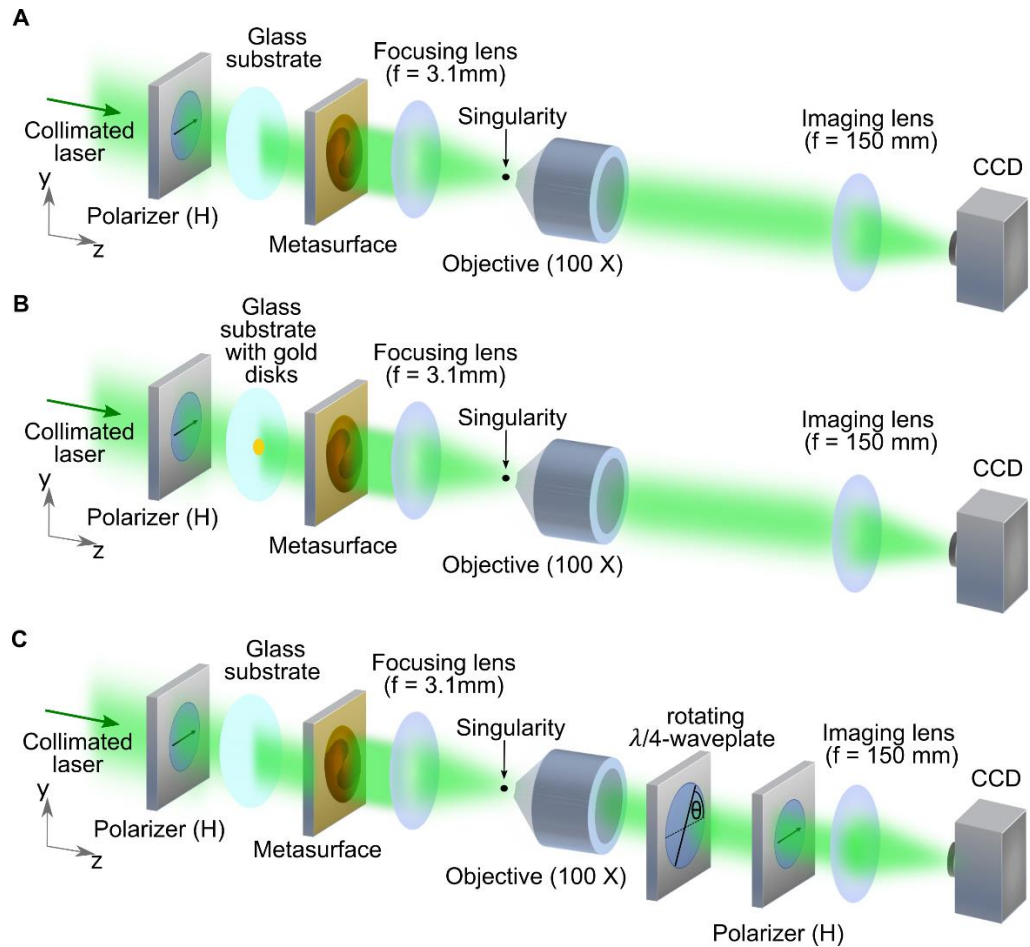

**Figure S13. Measurement setup. a,** Setup for intensity measurements. A collimated laser beam of adjustable wavelength is generated by a supercontinuum laser with a reconfigurable bandpass filter of 5 nm bandwidth and then passing through a horizontal polarizer and a glass substrate (to exclude effects from the glass substrate holding the perturbation gold disk in b)) before passing through the metasurface ($500 \mu m$). Light impinging outside of the metasurface area is blocked by a Al/Au mask. An aspheric lens of f=3.1mm (NA 0.08) is then used to focus the light and create the singularity. The singularity is then imaged with a microscope formed by an 100x objective (Nikon, NA=0.9), an imaging lens of f=15cm and a sCMOS camera (color sensor, pixel size $6.5\mu m \times 6.5\mu m$, dynamic range 21500:1) A motorized stage is used to move the objective along the z-direction. The metasurface, the glass substrate and the focusing lens are positioned on 3-axis stages with micrometer heads to enable precise positioning. **b,** Setup for perturbation measurements. The setup is identical to a) except that this time a gold disk on the glass substrate is moved in front of the metasurface to block part of the metasurface. **c,** Setup for polarization measurements. Measurement setup is identical to a), except that a quarter-waveplate (mounted on a motorized rotation stage) and a horizontal polarizer is added into the infinity space between the objective and the imaging lens.

## S5.1 Intensity measurement and data analysis

## S5.1.1 Intensity measurement and data analysis

A detailed setup description can be found in Figure S13 a. As the camera captures slices of the field in the xy plane, additional sweeps in wavelength and z position were performed in order to capture the field around the singularity in the 4D space (x,y,z,$\lambda$). The resolution of the sweep is: ($\Delta$x, $\Delta$y, $\Delta$z, $\Delta\lambda$)=(0.17$\mu m$, 0.17$\mu m$, 2$\mu m$, 2$nm$). Subsequently, the laser was turned off, 1000 background images were captured and the pixel-wise average was subtracted from the captured images to compensate for stray light from the room directly hitting the sCMOS. To compensate for laser power differences between different wavelengths, the captured images were normalized by the total pixel count (the area of capture is chosen large enough to capture all light passing from the laser through the system (Figure S13)). One remaining source of error is the finite bandwidth of the laser (5nm), that is larger than the wavelength steps of the measurement and hence increase the intensity at the singularity position.

Finding the singularity position in four dimensions:
To find the position of the singularity in the four-dimensional space, we loop through the positions in z and wavelength and search in each xy slice for the minimum intensity inside the circle of light (Figure S14 a). Due to the circular shape of the field surrounding the singularity, a weighted average of the image (excluding pixels smaller than the maximum pixel of the background image) gives a first estimate of the singularity position (Figure S14b). The minimum and its position can then be found by reducing the area of interest to an area inside of the light ball around the estimated position (Figure S14c). A repeating reduction in area of interest and updating of the estimation point then converges to the position of interest. The singularity position is determined by iterating these procedures for all position in z and wavelength, searching for the position $v = (x_0, y_0, z_0, \lambda_0)$ where the intensity of the point of interest within the circle of light is minimized.

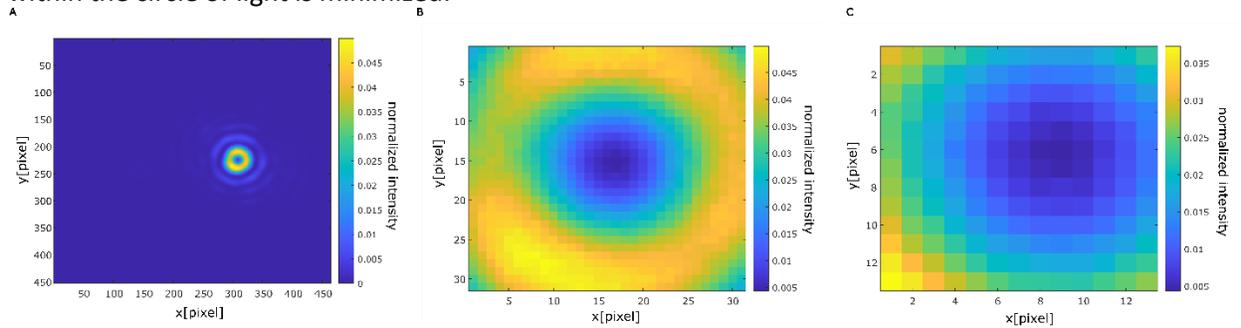

**Figure S14. Finding the singularity. a,** xy slice captured by the sCMOS camera, normalized by the total pixel count to compensate for wavelength dependent power changes. **b,** First reduced area of interest centred at the weighted average position. **c,** Reduced area of interest. The minimum field value is taken from this picture.

xy confinement (Figure 3c): The xy slice is plotted at $z = z_0$ and $\lambda = \lambda_0$ using the dB scale with $I_{dB} = 10\log_{10}(I/I_{max})$, where $I$ is the intensity of the xy slice and $I_{max}$ is the maximum intensity value in the four dimensional space of the captured data (after compensation for laser power difference between different wavelength). The dB scale is chosen, to better represent the range of fields close to the singularity.

z$\lambda$ confinement (Figure 3d). For each position in z and wavelength, the corresponding xy-slice is processed like Figure 3c) and the intensity of the minimal point within the circle of light is plotted.

xyz confinement (Figure 3e). The data was represented in dB, with $I_{max}$ being the maximum value in the whole four-dimensional dataset.

## S5.2 Perturbation measurement and data analysis

To experimentally demonstrate the perturbation protection of the singularity, we insert an opaque gold disk of diameter of 110 $\mu m$ in front of the metasurface, blocking part of the light from passing through. Due to the way the metasurface converts light into different polarizations over different areas of the metasurface, this corresponds to subtracting a polarized field in comparison to the unperturbed singularity. A detailed setup description can be found in Figure S13b. As the glass substrate holding the gold disk perturbation was added already in the unperturbed measurement (the gold mask was pushed out of the metasurface area), we can ensure that the perturbation effects are not caused by the glass substrate. The data analysis is described in S5.1.

## S5.3 Polarization measurement and data analysis

To analyze the polarization of the field around the singularity, we follow the mechanism described in (*11*). Adding a quarter waveplate and a horizontal polarizer to the infinity space between the objective and the imaging lens, one can retrieve the full Stokes vector at each pixel in the xy slice by rotating the quarter waveplate and capturing images at multiple angles $\theta$ (example measurement shown in Figure S15 left side):

$$A = \frac{2}{N}\sum_{n=1}^{N} I_n, \qquad B = \frac{4}{N}\sum_{n=1}^{N} I_n \sin 2\theta_n$$

$$C = \frac{4}{N}\sum_{n=1}^{N} I_n \cos(4\theta_n), \qquad D = \frac{4}{N}\sum_{n=1}^{N} I_n \sin 4\theta_n$$

Where $\theta_{n+1} - \theta_n = 180°/N$. The Stokes parameters then are determined by

$$S_0 = A - C, \quad S_1 = 2C, \quad S_2 = 2D, \quad S_3 = B$$

This procedure is repeated in a z region of $\pm 10\mu m$ around the singularity (stepsize $4\mu m$). Figure 3G in the main paper is created by evaluating the stokes vector on an elliptical surface of constant intensity around the singularity position with $\lambda = \lambda_0$ and polar plotting the corresponding polarization ellipse at position $(\rho, \phi)$ with

$$\rho = \text{atan2}\big((x - x_{\text{centre}}), (y - y_{\text{centre}})\big), \qquad \phi = |\text{atan}\,(\frac{\sqrt{(x - x_{\text{centre}})^2 + (y - y_{\text{centre}})^2}}{z})|$$

, where $(x_{\text{centre}}, y_{\text{centre}})$ is the singularity position. Each pixelated data ring of each xy slice (Figure S16b) is projected onto a perfect ring of radius $R = \frac{\sum_i^N \sqrt{(x_i)^2 + (y_i)^2}}{N}$ for representation reasons.

Figures S15 and S16 show other representations of the measured polarizations. Figure 15b shows a comparison between the experimental and simulated stokes vectors in the xy plane at $z = z_0$ and $\lambda = \lambda_0$, showing a good agreement. S16c,d show that the measured polarization states around the singularity cover the entire Poincare sphere, with an uneven distribution of datapoints due to the finite pixel size of the sCMOS camera (Figure S16 a,b).

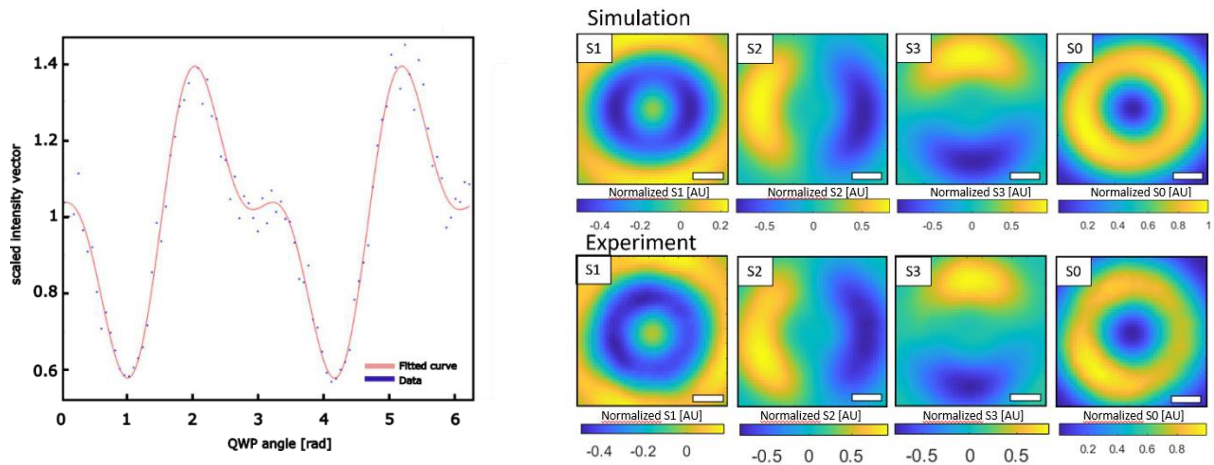

**Figure S15. Stokes vector extraction. a,** The intensity changes of a pixel depending on the rotation angle of the quarter waveplate (QWP) with respect to the horizontal polarizer. It can be used to extract the Stokes vector following the algorithm described in (*11*). **b,** Simulated (top) vs. experimental (bottom) stokes vector components. Scale bar 2um.

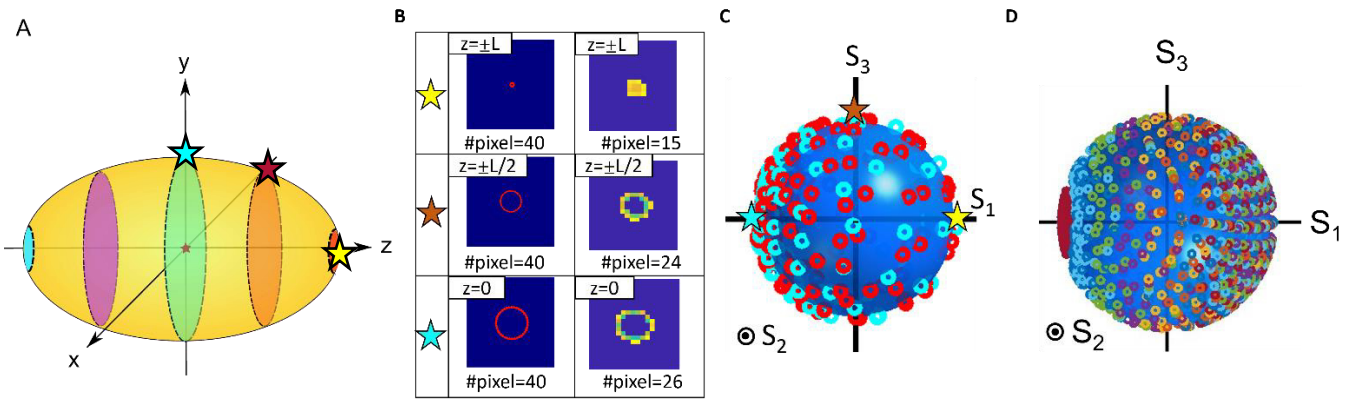

**Figure S16. Simulated and experimental polarization plotted on Poincare sphere. a,** Schematic of an surface around the singularity position (star in the origin) of equal intensity. xy-planes located at different z positions are marked in different colors, assuming the singularity is positioned at $(x,y,z,\lambda)=(0,0,0,\lambda_0)$. Stars mark different z positions for later comparison. **b,** Comparison of the number of datapoints available for different positions on the ellipsoid spheres for simulation (left) and experimental (right). The stars connect the position on the ellipsoid with the number of datapoints. While for the simulation the number of points on the ellipsoid is the same for each z position, the experimental data varies in datapoints due to the finite pixel size of the sCMOS camera. **c,** Due to the way the polarization is distributed on the ellipsoid (see Figure S3), many datapoints are available for $S_1=-1$ (as positioned on ellipsoid at z=0), but only few datapoints are available for $S_1=1$ (positioned on pole of ellipsoid at $z=\pm z_{max}$) causing a discrepancy of available datapoints between the left and right side of the Poincare sphere. Blue and red dots correspond to z>0 and z<0, respectively. **d,** Same as for c), with increased ellipsoid size. Different colors correspond to different z positions on the ellipsoid, showing that indeed different z positions on the ellipsoid correspond to different positions on the S1 axis on the Poincare sphere.